\documentclass[conference]{IEEEtran}

\usepackage[framed,autolinebreaks,useliterate]{mcode}
\usepackage{mathtools}
\usepackage{nicematrix}
\usepackage{float}
\usepackage{tabularx}
\usepackage{booktabs}
\usepackage{cancel}
\usepackage{enumitem} 
\usepackage{ytableau}
\usepackage{graphicx} 
\graphicspath{ {images/} }

\usepackage{color, colortbl}
\usepackage{multirow}

\usepackage{cuted}
\usepackage[linesnumbered,ruled,vlined]{algorithm2e} 

\usepackage[utf8]{inputenc}
\usepackage[english]{babel}
\usepackage{amsthm,amssymb}
\usepackage{amsmath}
\usepackage{bm}
\usepackage{optidef}
\usepackage{mathrsfs}

\usepackage[font=small,labelfont=bf,labelsep=space]{caption}
\usepackage{xcolor}
\def\BibTeX{{\rm B\kern-.05em{\sc i\kern-.025em b}\kern-.08em
    T\kern-.1667em\lower.7ex\hbox{E}\kern-.125emX}}

\usepackage{tikz}
\usetikzlibrary{calc,patterns,decorations.pathreplacing,decorations.pathmorphing,matrix,positioning,arrows.meta,fit,bending,backgrounds}
\theoremstyle{definition}
\newtheorem{example}{Example}
\newtheorem{theorem}{Theorem}
\newtheorem{proposition}{Proposition}
\newtheorem{lemma}{Lemma}

\newtheorem{definition}{Definition}

\newtheorem{remark}{Remark}

\DeclareMathOperator{\bin}{bin}
\DeclareMathOperator{\supp}{supp}

\DeclareMathOperator{\w}{w}
\DeclareMathOperator{\dist}{d}

\newcommand{\rank}{\mathrm{rank}}
\newcommand{\wm}{\w_{\min}}

\newcommand{\A}{\mathcal{A}}

\newcommand{\I}{\mathcal{I}}

\newcommand{\C}{\mathcal{C}}

\newcommand{\CI}{\mathcal{C}(\mathcal{I})}

\newcommand{\R}{\mathcal{R}}

\newcommand{\bA}{\mathbf{A}}
\newcommand{\bB}{\mathbf{B}}

\newcommand{\bve}{\bm{\varepsilon}}
\newcommand{\bvg}{\bm{\gamma}}

\newcommand{\bc}{\boldsymbol{c}}

\newcommand{\bv}{\boldsymbol{v}}

\newcommand{\bG}{\boldsymbol{G}}

\newcommand{\ind}{\operatorname{ind}}
\newcommand{\ev}{\operatorname{ev}}

\newcommand{\bi}{\bm{i}}
\newcommand{\idfg}{\mathbb{I}_{f=g}}

\newcommand{\ft}{\mathbb{F}_2}

\DeclareFontFamily{U}{mathx}{}
\DeclareFontShape{U}{mathx}{m}{n}{<-> mathx10}{}
\DeclareSymbolFont{mathx}{U}{mathx}{m}{n}
\DeclareMathAccent{\widecheck}{0}{mathx}{"71}

\newcommand{\Alow}{{\rm LTA}(m,2)}

\newcommand{\GL}{{\rm GL}(m,2)}
\newcommand{\Mon}{\mathcal{M}_{m}}
\newcommand{\weako}{\preceq_w}
\newcommand{\Rm}{{\mathbf {R}}_m}

\title{On Weight Enumeration and Structure Characterization of Polar Codes via Group Actions}
\author{
\IEEEauthorblockN{Vlad-Florin Drăgoi$^{\ddagger}$, and Mohammad Rowshan$^{\dagger}$}
\IEEEauthorblockA{$^\ddagger$Faculty of Exact Sciences, Aurel Vlaicu University, Arad, Romania\\
$^\dagger$School of Electrical Eng. and Telecom., University of New South Wales (UNSW), Sydney, Australia\\
vlad.dragoi@uav.ro, m.rowshan@unsw.edu.au }
}

\begin{document}

\maketitle
\pagestyle{plain}

\begin{abstract}
In this article, we provide a complete characterization of codewords in polar codes with weights less than twice the minimum distance, using the group action of the lower triangular affine (LTA) group. We derive a closed-form formula for the enumeration of such codewords. Furthermore, we introduce an enhanced partial order based on weight contributions, offering refined tools for code design. Our results extend previous work on Type II codewords to a full description of Type I codewords and offer new insights into the algebraic structure underlying decreasing monomial codes, including polar and Reed–Muller codes.
 \end{abstract}

\begin{IEEEkeywords}
Decreasing monomial codes, polar codes, weight distribution, closed-form formula, enumeration, partial order.
\end{IEEEkeywords}

\section{Introduction}
Polar codes \cite{arikan} were the first provably capacity-achieving codes constructed based on the polarization effect in the vector channel. The information-theoretic approach used for constructing polar codes results in poor weight distribution. Knowing the direct relation between weight distribution and error correction performance, to compensate for this weakness, pre-transformed polar codes such as CRC-polar and PAC codes \cite{arikan2,rowshan-pac1} are used to improve the weight distribution \cite{rowshan2023minimum}. For more details about various constructions of polar codes and pre-transformation, the interested readers are referred to \cite[Section VII]{rowshan2024channel}.
\subsection{Weight distribution of decreasing monomial codes}
\paragraph{Decreasing monomial codes -- permutation groups} Decreasing monomial codes \cite{bardet} are a large family of binary linear codes that includes well-known error-correcting codes, such as polar codes \cite{arikan} and Reed--Muller codes. A decreasing monomial code is a linear code that admits as basis the evaluation of a set of monomials in $m$ variables, say $\I$, on all the elements in $\ft^m.$ These monomials satisfy an order relation $\preceq$, which translates in part the reliability relations between the $2^m$ synthetic channels of polar codes. This algebraic description provided key insights into the structure of polar codes. Let us now recall key historical milestones relevant to our work. In \cite{bardet} properties such as duality, permutation group, and minimum weight codewords were revealed. Shortened and punctured decreasing monomial codes were studied from a cryptographic point of view in \cite{bardet2016crypt,Dragoi-Szocs}. The permutation group and its applications were one of the hottest topics in polar coding. The first subgroup of permutations, the lower triangular affine group $\Alow$, was identified in \cite{bardet} and further expanded to the block lower triangular affine group in \cite{GEEB21,li2021complete}. In \cite{li2021complete} the authors demonstrated that these are the only affine automorphisms of polar codes. In \cite{IU22} it was shown that in the asymptotic regime the permutation group of polar codes converges to $\Alow.$ However, decreasing monomial codes admit other permutations than affine transformations \cite{ma2024-permutations}. One of the most notable practical contributions of the permutation group was in parallelized decoding \cite{GEECB21,  PBL21, IU22}.   
\begin{figure*}[ht]
\centering
\resizebox{\textwidth}{!}{
\begin{tikzpicture}
\draw (-5,1.75) node {$\ev(x_1x_2x_3)$}
        (-3,0.25) node {$\ev(x_1x_4x_5)$}
         (1.5,1) node {\small$\ev(x_1x_2x_3x_4x_5)$};
\draw  (-4,2) rectangle (-1.2,1.5);
\pattern[pattern=north east lines, pattern color=blue] (-4,2) --(-4,1.5) --(-1.9,1.5)--(-1.9,2)--cycle;
\draw[dashed] (-1.9,1.25) -- (-1.9,2.25);
\pattern[pattern=north east lines,  pattern color=red] (-1.9,1.5) --(-1.2,1.5) --(-1.2,2)--(-1.9,2)--cycle;
 \draw[latex-latex] (-1.9,2.5) -- node[above] (O1) {\footnotesize$\frac{1}{4}2^{6-3}$} (-1.2,2.5);
 \draw[latex-latex] (-4,2.5) -- node[above] (O2) {\footnotesize$\frac{3}{4}2^{6-3}+$} (-1.2,2.5);
\draw  (-1.9,0) rectangle (-1.9+2.8,0.5);
\pattern[pattern=north east lines, pattern color=blue] (-1.2,0) --(-1.2,0.5) --(-1.9+2.8,0.5)--(-1.9+2.8,0)--cycle;
\draw[dashed] (-1.2,-0.25) -- (-1.2,0.75);
\pattern[pattern=north east lines,  pattern color=red] (-1.2,0.5) --(-1.9,0.5) --(-1.9,0)--(-1.2,0)--cycle;
\draw[-{Stealth[length=5pt]},thick] (-1.6,1.75) to [bend right] (0,1);
\draw[-{Stealth[length=5pt]},thick] (-1.6,0.25) to [bend left] (0,1);
 \draw[latex-latex] (-4,-0.4) -- node[below] (O2) {\footnotesize$\frac{3}{4}2^{6-3}+\frac{3}{4}2^{6-3}=1.5\times 2^{6-3}$} (0.9,-0.4);
  \draw (-2,-1.5) node {\bf a) $1.5\wm$ Type II};

  \draw (-5+10,1.75) node {$\ev(x_1x_2x_3)$}
        (-3+9.5,0.25) node {$\ev((x_1+x_0)x_4x_5)$}
         (1.5+10.5,1) node {\small$\ev((x_0+1)x_1x_2x_3x_4x_5)$};
\draw  (-4+10,2) rectangle (-1.2+10,1.5);
\pattern[pattern=north east lines, pattern color=blue] (-4+10,2) --(-4+10,1.5) --(-1.9+10.25,1.5)--(-1.9+10.25,2)--cycle;
\draw[dashed] (-1.9+10.25,1.25) -- (-1.9+10.25,2.25);
\pattern[pattern=north east lines,  pattern color=red] (-1.9+10.25,1.5) --(-1.2+10,1.5) --(-1.2+10,2)--(-1.9+10.25,2)--cycle;
 \draw[latex-latex] (-1.9+10.25,2.5) -- node[above] (O1) {\footnotesize$\frac{1}{8}2^{6-3}$} (-1.2+10,2.5);
 \draw[latex-latex] (-4+10,2.5) -- node[above] (O2) {\footnotesize$\frac{7}{8}2^{6-3}+$} (-1.2+10,2.5);
\draw  (-1.9+10.25,0) rectangle (-1.9+2.8+10,0.5);
\pattern[pattern=north east lines, pattern color=blue] (-1.2+10,0) --(-1.2+10,0.5) --(-1.9+2.8+10,0.5)--(-1.9+2.8+10,0)--cycle;
\draw[dashed] (-1.2+10,-0.25) -- (-1.2+10,0.75);
\pattern[pattern=north east lines,  pattern color=red] (-1.2+10,0.5) --(-1.9+10.25,0.5) --(-1.9+10.25,0)--(-1.2+10,0)--cycle;
\draw[-{Stealth[length=5pt]},thick] (-1.6+10.1,1.75) to [bend right] (0+10,1);
\draw[-{Stealth[length=5pt]},thick] (-1.6+10.1,0.25) to [bend left] (0+10,1);
 \draw[latex-latex] (-4+10,-0.4) -- node[below] (O2) {\footnotesize$\frac{7}{8}2^{6-3}+\frac{7}{8}2^{6-3}=1.75\times 2^{6-3}$} (0.9+10,-0.4);
  \draw (-2+10,-1.5) node {\bf a) $1.75\wm$ Type I};
 \end{tikzpicture}
 }
\caption{Two distinct weights in the Minkowski sum of $\Alow\cdot x_1x_2x_3+\Alow\cdot x_1x_4x_5$, for $m=6.$ a) $\mu=2$ or equivalently $1.5\wm$ Type II codeword (same as \cite{vlad1.5d}), and b) $\mu=3$ or $1.75\wm$ Type I codeword. The product $x_1x_2x_3(x_1+x_0)x_4x_5$ was refactored into $(x_0+1)x_1x_2x_3x_4x_5$ which produces a vector of weight $1.$}\label{fig:example1_intro}
\end{figure*}
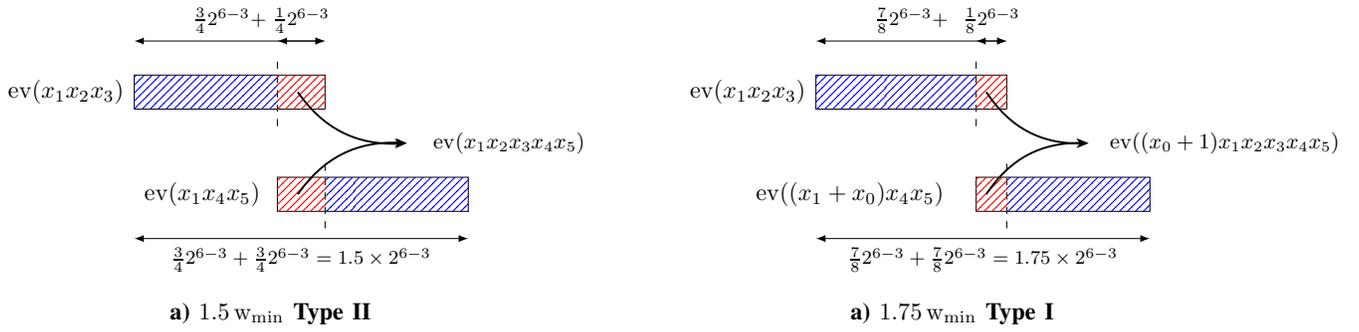

\paragraph{Weight enumeration} Recently, a strong focus has been directed towards weight enumeration for polar codes. Although algorithmic methods are among the most widespread solutions \cite{liu_analys,valipour,zhang_prob,fossorier,polyanskaya,yao}, they suffer from several drawbacks, the most significant being their high complexity (weight enumeration is possible only for codes with lengths up to $n=128$). That is why closed formulae were proposed. The formation of minimum weight codes in terms of rows of the generator matrix of polar codes was discussed in \cite{rowshan2023formation}. In \cite{bardet} based on the action of $\Alow$ a formula for the number of minimum weight codewords was proposed. This was further extended to $1.5\wm$-weight codewords in \cite{vlad1.5d}. The main difficulty between $\wm$ and the higher weights is in characterization of Minkowski sums of orbits. Indeed, a minimum weight codeword lies inside one orbit $\Alow\cdot f$ where $f$ is a maximum degree monomial. However, higher weights are formed by summing several minimum weight codewords, making the characterization and counting more challenging. The result on $1.5\wm$ was further extended to weight up to $2\wm$, for Type II codewords in \cite{rowshan2024weight}. In parallel, Type II and Type I codewords were enumerated in \cite{Ye2024-weightdistrib}, based on the extension of Kasami and Tokura's theorem for Reed--Muller codes \cite{kasami1970weight}. Such results are in line with older contributions on the special case of Reed-Muller codes \cite{kasami1970weight,kasami1976w2.5d}. 

\subsection{Contributions}
In this study, we extend the findings presented in \cite{vlad1.5d} and \cite{rowshan2024weight} to characterize codewords of Type I. We introduce a new subgroup of the $\Alow$ in order to characterize special cases of sums of orbits $\Alow\cdot f+\Alow\cdot g$ where $f,g$ are maximum degree monomials with $\deg(\gcd(f,g))>r-\mu.$ Since our formulae are based on the action of $\Alow$ they are different from \cite{Ye2024-weightdistrib}. We provide numerical examples that align with and go beyond known examples, such as polar, Reed--Muller and RMxPolar codes. In addition, we provide arguments on how to use our characterization to improve recent partial orders based on weight contribution \cite{rowshan2024weight}. Let us detail our contributions.

\paragraph{Starting point}
To completely characterize Type I codewords we build upon a recent perspective codewords in terms of $\Alow$ action. Given a decreasing monomial code $\CI$ with $r=\max_{f\in\I}\deg(f)$, we start from \cite{rowshan2024weight} (Theorem 2) which states that any such codeword $\ev(P)$ ($P$ being a polynomial in the span of the monomial basis $P\in\mathrm{span}(I)$) is of the form $P=y_1\dots y_{r-\mu}(y_{r-\mu+1}\dots y_{r}+y_{r+1}\dots y_{r+\mu})$ with $P\in \Alow\cdot f+\Alow\cdot g$ with $f,g\in \Mon$ having $\deg(f)=\deg(g)=r.$ and $y_i$ being independent linear forms. The weight of $\ev(P)$ equals $\w_{\mu}=2^{m+1-r}-2^{m+1-r-\mu}$ and the parameter $\mu$ satisfies $m\geq r+\mu,r\geq \mu\geq 3.$ 

A crucial tool in our proofs is the independence condition for linear forms. Already mentioned in \cite{vlad1.5d,rowshan2024weight}, it states that any finite product of linear independent forms $l_i$ can be rewritten such that maximum variables in the new linear forms $l_i^{*}$ are distinct. A direct consequence of this property is that any Type I codeword should have at least $r+\mu$ distinct variables in $y_1\dots,y_{r+\mu}.$ This observation plays a central role throughout our work.   

Another useful tool is the number of free variables taken into consideration when applying $\Alow$ on a monomial $f=x_{i_{1}}\dots x_{i_{r}}$ with $i_1<\dots< i_r.$ More precisely, we will use the fact that the number of free variables taken together with $x_i\in\ind(f)$ is $\lambda_f(x_i)=|J_f(i)|,$ where $J_f(i)=\{j\in[0,i) \mid j\notin\ind(f)\}.$ This was previously used in \cite{bardet} for enumerating minimum weight codewords, where $\lambda_f=(i_{r}-(r-1),\dots,i_1-0)$ and $|\lambda_f(f)|=\sum_{i\in\ind(f)}\lambda_f(x_i)$, (or simply $\lambda_f$) was defined. We shall employ these objects several times in our paper. 

\paragraph{Type I codewords - conditions on monomials} 
To completely characterize Type I codewords of Decreasing monomial codes we first analyze the exiting conditions on monomials $f,g$ that define the orbits for $P=\Alow\cdot f+\Alow\cdot g.$ We will denote by $\I_r$ the set of monomials in $\I$ with degree $r.$ We demonstrate that a pair $(f,g)$ can define a valid Type I generating codewords if the following conditions are met 
\begin{itemize}
\item both have maximum degree, i.e., $\deg(f)=\deg(g)=r$ ($f,g\in\I_r$);
\item they share at least $r-\mu$ coomon variables. In other words their common divisor $h=\gcd(f,g)$ satisfies $\deg(h)\geq r-\mu.$ Three distinct cases follow 
\begin{itemize}
    \item \textbf{Case A1.} if $f$ and $g$ share $r-\mu$ common variable then $(f/h,g/h)$ are co-prime and $(f,g)$ is a valid pair
    \item \textbf{Case A2.} if $f$ and $g$ share more than $r-\mu$ common variable, say $r-\mu+l.$ Then, we need to check whether there are $l$ variables among the $r-\mu+l$, say $x_{i_1}\dots x_{i_l}$ such that a shift on each variable $x_{i_1-\epsilon_1}\dots x_{i_l-\epsilon_l}$ produces a degree $l$ monomial that divides the complement of $fg$ (the monomial $x_0\dots x_{m-1}/(fg)$) (see Figure \ref{fig:example1_intro} for $f=x_1x_2x_3, g=x_1x_4x_5$).
    \item \textbf{Case B1.} if $f=g$ and $f\not\in\I$ then since $f$ is not a term in $\Alow\cdot f+\Alow\cdot f$ this could eventually produce codewords of $\CI$. For that $f$ needs to admit a factor of degree $r-\mu$ denoted $h$ and a shift condition as in the previous case could be applied on $f/h.$ Also we have degree $r$ monomials among the terms of Minkowski sum of orbits that belong to $\I.$
\end{itemize}
\end{itemize}

The technicalities involved in our proofs also include the definition of a new order relation $\prec_{sh}$ that is closely related to the previous and well-known order $\preceq_{sh}$ valid on monomials with identical degrees. To be more exact, recall that for $\deg(f)=\deg(g)=s, f\preceq_{sh} g$ if $\forall\;1\leq\ell\leq s\;\text{ we have }\;  i_\ell \le j_\ell$, where $f=x_{i_1}\dots x_{i_s}$, $g=x_{j_1}\dots x_{j_s}$. Now, $f\prec_{sh} g$ if $ i_\ell <j_\ell$ for all values of $\ell.$

\begin{table*}[]
    \centering
    \begin{tabular}{c|c|c|c}
    \toprule
        \multicolumn{2}{c|}{Type} & Conditions & Cardinality of orbit \\
        \midrule
        \multicolumn{2}{c|}{Type I} &  $1\leq i\leq \mu,\;f_i\in \I_r,\quad  h=\gcd(f_i,f_j) \in \I_{r-2}$ & $\dfrac{2^{r-2+2\mu+ |\lambda_h|+\sum\limits_{i=1}^{\mu}|\lambda_{f_i}(\frac{f_i}{h})|}}{2^{\sum\limits_{(f_i,f_j)}\alpha_{\frac{f_i}{h},\frac{f_j}{h}}}}$\\
        \midrule
        \multirow{10}{*}{Type II} & \multirow{2}{*}{Sub-type A.1} & $f,g\in \I_r,\; h=\gcd(f,g),\;\deg(h)=r-\mu$& $2^{r+\mu+|\lambda_h|+\left|\lambda_f\left(\frac{f}{h}\right)\right|+\left|\lambda_g\left(\frac{g}{h}\right)\right|}$\\
        &&&\\
        \cline{2-4}
        &\multirow{4}{*}{Sub-type A.2}&&\multirow{4}{*}{${2^{r+\mu-\idfg+|\lambda_h|+\left|\lambda_f\left(\frac{f}{h}\right)\right|+\left|\lambda_g\left(\frac{g}{h}\right)\right|}}\prod\limits_{j=1}^{\deg(h^{*})}\left(1-\frac{1}{2^{|J_{fg}{(i_j)}|-(j-1)}}\right)$}\\
         & & $f,g\in \I_r,\;  \gcd(f,g)=hh^{*},\; \deg(h)=r-\mu,$ &\\
         && $h_s^{*}\preceq_w \widecheck{fg},\;h_s^{*}\prec_{sh} h^{*}$ &\\
         &&&\\
         \cline{2-4}
         &\multirow{4}{*}{Sub-type B.1}&&\multirow{4}{*}{${2^{r+\mu-1+|\lambda_h|+\left|\lambda_f\left(\frac{f}{h}\right)\right|}}
            \prod\limits_{j=1}^{\deg(h^{*})}\left({2^{|J_f^{\I}{(i_j)}|}-2^{j-1}}\right)$}\\
         & & $f\not\in \I\,;  \deg(f)=r,\; f=hh^{*},\; \deg(h)=r-\mu, $& \\
         &&$h_s^{*}\preceq_w \check{f},\; h_s^{*}\prec_{sh} \frac{f}{h}$&\\
         &&&\\
         \bottomrule
    \end{tabular}
    \caption{Counting formulae for $\w_{\mu}$-weight codewords of decreasing monomial codes}
    \label{tab:formula-all-intro}
\end{table*}

\paragraph{Type I codewords - Minkowski sums of orbits}
Characterizing the Minkowski sum of orbits $\Alow\cdot f+\Alow\cdot g$ was a complex and challenging task for all three cases. In \cite{vlad1.5d} the authors introduced the notion of collisions in order to determine the cardinality of the Minkowski sum. The difficulty comes from the fact that all the structure that $\Alow\cdot f$ inherits from the group properties cannot be transferred to the Minkowski sum. Therefore, each case had to be considered apart. Another difficulty consists in breaking down the elements inside a Minkowski sum into clusters with respect to the Hamming weight weight. Take for example $f=x_1x_2x_3,g=x_1x_4x_5$ provided in Figure \ref{fig:example1_intro}. We see that in $\Alow\cdot f+\Alow\cdot g$ we have at least two different weights, i.e., $1.5\wm$ for example the evaluation of $x_1(x_2x_3+x_4x_5)$, and $1.75\wm$ for example the evaluation of $x_1x_2x_3+(x_1+x_0)x_4x_5$. And there is also, $2\wm$, the evaluation of $x_1x_2x_3+(x_1+1)x_2x_3.$ 
 
For \textbf{Case A1.} we showed that there are no collisions and thus, the cardinality of $\Alow\cdot h\left(\Alow_f\cdot f/h+\Alow_g\cdot g/h\right)$ is the product of the cardinality of each orbit. We have built our proof on a more general properties. One of these properties state that if the degree of $\gcd(f,g)$ is strictly smaller than $r-2$ then the cardinality of the Minkwoski sum is the sum of the cardinality of the two orbits.

Regarding \textbf{Case A2.} the situation becomes more complex. Suppose we have $f,g\in\I_r$ with $h\in\I_{r-\mu}$ such that $\gcd(f,g)=hh^{*}$ with $\deg(h^{*})\geq 1.$ The $\w_{\mu}$-weight codewords are evaluation of polynomials that are composed of $r+\mu$ independent linear forms lying inside the orbit $\Alow_h\cdot h\left(\Alow_f\cdot \frac{f}{h}+\Alow_g \cdot \frac{g}{h}\right).$ While $\Alow_h\cdot h \times \Alow_f\cdot \frac{f}{h}$ is a product of $r$ independent linear forms, the remaining part might not provide the extra $\mu$ independent linear forms, due to the existence of $h^{*}$ on both $f$ and $g.$ Hence, the group action on $h^{*}$ should be restricted to only those matrices $\bB$ that produce independent linear forms with the preexisting forms coming from the action on $f.$ Here we introduce a new sub-group $\Alow_f^{g}$ that acts on $f/h$ group that is formed by the matrices with the aforementioned condition. In other words, if on $g$ we allow the complete group action on the $h^{*}$, then on $f$ we need to split the action into two distinct parts, a free part on $\frac{f}{\gcd(f,g)}$ and a restricted part on $h^{*}.$ 
We also determine a closed formula for the cardinality of this new orbit. We do connect our formula with the preexisting Young diagrams for counting the elements in $\Alow\cdot f$ (see \cite{bardet,dragoi17thesis}). With this new group action the cardinality of $\Alow\cdot h\left(\Alow_f\cdot f/h+\Alow_g\cdot g/h\right)$ equals $ \left|\Alow_h\cdot h\right|\left|\Alow_f\cdot \frac{f}{h}\right|\left|\Alow_g^{{f}{}} \cdot \frac{g}{h}\right|.$ 

The \textbf{Case B1.} is similar to the previous case. However, since $f\notin\I$ we need to take into account the monomials od degree $r$ produced by $\Alow\cdot f$ that are not in $\I.$ Hence, we introduce a new subgroup of $\Alow$, denoted $\Alow_f^{f,\I}$ that formally describes all the independent variables one can consider in each $x_i$ from $f$, such that all the monomials in $\Alow_f^{f,\I}\cdot f/h+\Alow_f\cdot f/h$ belong to $\I.$ The set $J_f^{\I}(i)=\{j\in[0,i-1) \mid j\not\in\ind(f),x_jf/x_i\in\I\}$ will be required for this new group action. We further provide a closed formula for the cardinality of $\Alow_f^{f,\I}.$

\paragraph{Applications: counting formulae and weight contribution partial orders}
The first application we propose is a closed formulae for the number of Type I $\w_{\mu}$ codewords of decreasing monomial codes. We demonstrate that orbits do not intersect and hence, we complete the table of formulae for estimating codewords with weights between $\wm$ and $2\wm$ (see Table \ref{tab:formula-all-intro}). Next, we give a potential use of our characterization for improving the partial order proposed in \cite{rowshan2024weight}. In \cite{rowshan2024weight} the authors used $\wm$ to quantify the impact of a monomial. Since there are non-comparable monomials with respect to reliability and weight-contribution we can use Type I structure to determine whether the impact of monomials on higher than $\wm$ weights.

A short version of this article was accepted for publication at ISIT 2025. For a smoother reading experience we have decided to include all of our proofs in the Appendix.

\section{The algebraic formalism behind polar and Reed-Muller codes}
The majority of our notations and definitions for the algebraic formalism as well as are taken from \cite{bardet,dragoi17thesis,rowshan2024weight,vlad1.5d}. For generic coding theory we consider \cite{lin_costello} while polar coding definitions and notations come from \cite{arikan}. 

\begin{table*}[ht]
    \centering{
\begin{tabular}{cccc||c c c c c c c c}
Index - $i$&$\bin(i)$&$\bin(2^m-1-i)$&$g$ & 111&011&101&001&
110& 010& 100& 000\\  
\hline
\hline
0&(000)&(111)&$\ev(x_0x_1x_2)$&1&0&0&0& 0&0&0&0\\
1&(100)&(011)&$\ev(x_1x_2)$&1&1&0&0&0&0&0&0\\
2&(010)&(101)&$\ev(x_0x_2)$&1&0&1&0&0&0&0&0\\
3&(110)&(001)&$\ev(x_2)$&1&1&1&1& 0&0&0&0\\
4&(001)&(110)&$\ev(x_0x_1)$ &1&0&0&0&1&0&0&0\\
5&(101)&(010)&$\ev(x_1)$&1&1&0&0&1&1&0&0\\
6&(011)&(100)&$\ev(x_0)$&1&0&1&0&1&0&1&0\\
7&(111)&(000)&$\ev(1)$&1&1&1&1& 1&1&1&1\\
\end{tabular}
}
    \caption{The matrix $\bG_{2^3}$ as evaluation of monomials in $\Mon$. }
    \label{tab:my_label}
\end{table*}

 \subsection{Basic Concepts in Coding Theory and Notations}
 \label{subsec:basic}

We denote by $\ft$ the finite field with two elements and by $+$ the addition operator in $\ft$. The symmetric group of order $N$ is denoted by $\mathrm{S}_N.$
Also, subsets of consecutive integers are denoted by $[\ell,u]\triangleq\{\ell,\ell+1,\ldots,u\}$. 
{\color{black}The binary expansion of an integer $i$ is denoted by $\bin(i)=(i_0,i_1,\dots,i_{m-1})\in \ft^m$, where $i_{m-1}$ is the most significant bit, i.e., $i=\sum_{j=0}^{m-1}i_j 2^{j}.$} We will use an order relation on the set $\ft^m$ by considering the decreasing index order on $[0,2^m-1]$. For example the elements in $\ft^2$ will be ordered as follows $\{(1,1),(0,1),(1,0),(0,0)\}$ and it corresponds to $\{3,2,1,0\}.$

The \emph{support} of a vector $\bc = (c_0,\ldots,c_{N-1}) \in \ft^N$ is defined by $\supp(\bc) \triangleq \{i \in [0,N-1] \mid c_i \neq 0\}$. 
The cardinality of a set is denoted by $|\cdot|.$ and the set difference by $\backslash$. The complement of a support is $\supp(\bc)^c=[0,N-1]\setminus \supp(\bc).$ The Hamming \emph{weight} of $\bc \in \ft^N$ is $\w(\bc)\triangleq |\supp(\bc)|$. 
Given two vectors $\bc = (c_0, c_1, \ldots, c_{N-1})$ and $\bc' = (c'_0, c'_1, \ldots, c'_{N-1})$, 
the Hamming \emph{distance} between $\bc$ and $\bc'$ is defined by $\dist(\bc,\bc') = |\{i \in [0,N-1] \mid c_i \ne c'_i\}|$. 

A $K$-dimensional subspace $\C$ of $\ft^N$ is called a linear $(N,K)$ \emph{code} over $\ft$ ($N$ is the length, $K$ the dimension and $R\triangleq K/N$ is the code rate). $\C$ is said to be a linear $(N,K,d)$ code if $\C$ is a linear $(N,K)$ code and
its minimum distance, 
\[
d_{\mathrm{min}} = \dist(\C) \triangleq \min_{\bc,\bc' \in \C, \bc \neq \bc'} \dist(\bc,\bc')
\]
is $d.$ The Hamming weight induces the Hamming distance and vice-versa \cite[Section 3.3]{lin_costello}, hence 
 \[\wm\triangleq\min_{\bc\in\C,\bc\neq 0}\w(\bc)=\dist(\C).\] 

The vectors in $\C$ are called \emph{codewords} and can be collected with respect to their weight $\w$ in 
$
    W_{\w}(\C) = \{ \bc\in\C \mid \w(\bc)=\w \}.
$
Moreover, for any $(N,K)$ one can define the \emph{weight enumerator polynomial} \[W(\C;x)=\sum_{\w=0}^{N}|W_{\w}(\C)|x^{\w}.\]
A \emph{generator matrix} $\bG$ of an $(N,K)$ code $\C$ is a $K \times N$ matrix in $\ft^{K\times N}$ whose rows are $\ft$-linearly independent codewords of $\C$. Then $\C = \{\bv \bG \colon \bv \in \ft^K\}$.

\subsection{Multivariate monomials and polynomials over $\ft$}

Let $m$ be a fixed integer, which will represents the number of different variables $\mathbf{x}\triangleq(x_{0},\dots,x_{m-1})$. 
Let $\ft[x_0,\dots{},x_{m-1}]$ be the set of polynomials in $m$ variables. Since we are dealing with binary codes, we will identify $x_i$ with $x_i^2$ (using the Frobenius endomorphism) and consider the ring $\Rm=\ft[x_0,\dots{},x_{m-1}]/(x_0^2-x_0,\dots,x_{m-1}^2-x_{m-1})$.

A monomial in $\Rm$ can be defined as
$$\mathbf{x}^{\mathbf{i}}=\prod_{j=0}^{m-1}x_{j}^{i_{j}}=x_{0}^{i_{0}} \cdots x_{m-1}^{i_{m-1}},$$ where $\mathbf{i}=\bin(i)=(i_{0},\dots,i_{m-1})$ with $i_j\in\{0,1\}.$ Let  
$
\Mon \triangleq \left\{\mathbf{x}^{\mathbf{i}} \mid\mathbf{i} \in \mathbb{F}_{2}^{m}\right\}
$ denote the set of all monomials in $\Rm.$
We have thus defined a one-to-one mapping between any integer $i\in[0,2^m-1]$ and a monomial $\mathbf{x}^{{\bin(i)}}.$ 
Since for $\ft^m$ we have used the decreasing index order we shall still with this convention for monomials as well

\[
\begin{array}[h]{ccccc}
[0,2^m-1]    & \to &\Mon\\
i& \mapsto &\mathbf{x}^{{\bin(2^m-1-i)}}
\end{array}
\]

For example, for $m=2$ we have $\Mon=\{x_0x_1,x_1,x_0,\bm{1}\}.$ 

Further we will define the \emph{support of monomial} $f=x_{l_1}\dots x_{l_s}$ by $\ind(f)=\{l_1,\dots,l_s\}$. The \emph{multiplicative complement} of a monomial $f$ is $\check{f}=\frac{x_0\dots x_{m-1}}{f}.$ In other words, we have $\ind(f)\cup \ind(\check{f})=[0,m-1]$ and $\ind(f)\cap\ind(\check{f})=\emptyset.$ Also, the degree of a monomial is $\deg(f)=|\ind(f)|.$ The degree induces a ranking on any monomial set $\I\subseteq\Mon$, i.e., $\I=\bigcup_{j=0}^{m}\I_j$, where $\I_j=\{f\in\I\mid\deg(f)=j\}.$ We denote the indicator function $\idfg$, which equals $0$ whenever $f\neq g$ and $1$ for $f=g.$

\subsection{Decreasing Monomial Codes}
Next we will define the evaluation function that associates to a polynomial $g \in \Rm$ the binary vector denoted by 
$\ev(g)$ in $\ft^{2^m}.$ 

\begin{definition}
Let $\ft^m$ ordered w.r.t. the decreasing index order. For $g\in \Rm$ define the evaluation function
\[
\begin{array}[h]{ccccc}
\Rm    & \to &\ft^{2^m}\\
g& \mapsto &\ev(g) = \big(g(\bi) \big)_{\bi \in \ft^m}
\end{array}
\]
\end{definition}

The function $\ev(\cdot)$ defines a vector space isomorphism between the vector space $(\Rm,+,\cdot)$ and $(\ft^{2^m},+,\cdot).$ We can now define monomial/polynomial codes. 
\begin{definition}[\cite{dragoi17thesis}]
Let $\I\subseteq\Rm$ be a finite set of polynomials in $m$ variables. 
The linear code  defined by $\I$ is the vector subspace $\C(\I) \subseteq \ft^{2^m}$ 
generated by $\{ \ev(f) ~|~ f \in \I\}$.

\begin{itemize}
 \item When $\I\subseteq \Rm$ we say that $\C(\I)$ is a \emph{polynomial code}.

\item When $\I\subseteq \Mon$ we say that $\C(\I)$ is a \emph{monomial code}.
\end{itemize}
\end{definition}

Now let us see how to construct monomial codes. For that we shall define the Kronecker product matrix
\[
\bG_{2^m} \triangleq \underbrace{\begin{pmatrix} 1 & 0 \\ 1 & 1 \end{pmatrix} \otimes \cdots 
\otimes \begin{pmatrix} 1 & 0 \\ 1 & 1 \end{pmatrix}}_{m \;\text{ times}}.
\]

In \cite{dragoi17thesis} it was demonstrated that $\bG_{2^m}$ is the monomial evaluation basis for the vector space $\ft^{2^m}$, fact that relies on the following diagram

\[
\begin{array}[h]{ccccc}
[0,2^m-1]    & \to &\Mon   & \to &\ft^{2^m}\\
i& \mapsto &g\triangleq\mathbf{x}^{{\bin(2^m-1-i)}} & \mapsto &\bG_{2^m}[i]=\ev(g) 
\end{array}
\]

We shall further continue with the definition of Decreasing monomial codes. Let ``$|$" denote divisibility between monomials, i.e., $f|g$ iff $\ind(f)\subseteq\ind(g).$ Also, the greatest common divisor of two monomials is $\gcd(f,g)=h$ with $\ind(h)=\ind(f)\cap\ind(g).$

\begin{definition}[\cite{bardet}]\label{def:order}Let $m$ be a positive integer and $f,g\in\Mon.$ Then $f\weako g$ if and only if $ f|g.$ When $\deg(f)=\deg(g)=s$ we say that $f\preceq_{sh} g$ if $\forall\;1\leq\ell\leq s\;\text{ we have }\;  i_\ell \le j_\ell$, where $f=x_{i_1}\dots x_{i_s}$, $g=x_{j_1}\dots x_{j_s}$. 
Define $f\preceq g\quad \text{iff}\quad \exists g^*\in \Mon\;\text{s.t.}\; f\preceq_{sh} g^*\weako g$.
\end{definition}

Here we will need an order relation stricter than $\preceq_{sh}.$ 
\begin{definition}
    Let $f,g\in\Mon$ with $f=x_{i_1}\dots x_{i_s}$, $g=x_{j_1}\dots x_{j_s}.$ We define
    \[f\prec_{sh} g \Leftrightarrow \forall\;1\leq\ell\leq s\;\text{ we have }\;  i_\ell < j_\ell.\]
\end{definition}
Notice that $f\prec_{sh}g\Rightarrow f\preceq_{sh}g$, while the converse is not valid. For example $x_0x_2\preceq_{sh}x_0x_3$ but $x_0x_2\not\prec_{sh}x_0x_3.$ There are particular cases when the converse is also true. 
\begin{lemma}\label{lem:sh-strict-sh}
    Let $f,g\in\Mon.$ The following holds
    \begin{itemize}
        \item if $\gcd(f,g)=1$ then $f\preceq_{sh}g\Leftrightarrow f\prec_{sh} g.$
        \item if $h=\gcd(f,g)\neq 1$ then $f\preceq_{sh} g\Leftrightarrow f/h\prec_{sh} g/h.$
    \end{itemize} 
\end{lemma}
The proof of this lemma is rather straightforward from the definition of $\prec_{sh}$ and $\preceq_{sh}.$
    
Notice that we have $x_0\prec_{sh} x_1\prec_{sh} \dots \prec_{sh} x_{m-1}.$ 
Understanding decreasing sets is crucial since they determine the structure of decreasing monomial codes. We can now recall the definition of monomial decreasing sets.

\begin{definition}[\cite{dragoi17thesis,bardet}]
      A set $\I \subseteq \Mon$ is \emph{decreasing}  if and only if ($f \in \I$ and $g \preceq f $) implies $g \in \I$. 

      A decreasing closed interval with respect to $\preceq$ is $[f,g]_{\preceq}=\{h\in \Mon \mid  f\preceq h\preceq g\}.$
\end{definition}

Any monomial code $\C(\I)$ with $\I$ decreasing is called decreasing monomial code. Both Polar and Reed-Muller codes are decreasing monomial codes \cite{bardet}. For $\R(r,m)$ we have
\begin{equation}
\label{eq:RM_decreasing}
 \R(r,m)=\C\left([1, x_{m-r}\cdots{} x_{m-1}]_{\preceq}\right).
\end{equation}

\subsection{Permutation group of Decreasing Monomial Codes}

The set of applications that leave a code $\C$ globally invariant, forms the \emph{automorphism group} of the code $\C$, which is denoted by $\operatorname{Aut}(\C)$. In the case of binary codes $\operatorname{Aut}(\C)\subset \mathrm{S}_N$, thus being all permutations. 

Decreasing monomial codes admit as subgroup of permutations the lower triangular affine group, $\Alow.$ An affine transformation over $\ft^m$ is represented by a pair $(\bB, \bve)$ where $\bB=(b_{i,j})$ is an invertible matrix lying in the general linear group $\GL$ and $\bve$ in $\ft^m$. Lower triangular affine transformation are defined by lower triangular binary matrices $\bB\in\GL$ with $b_{i,i}=1$ and $b_{i,j}=0$ whenever $j>i$. More on this group action on decreasing monomial codes can be found in \cite{dragoi17thesis,vlad1.5d}.

Let us recall how $\Alow$ acts on a monomial, say  $g=\prod_{i\in\ind(g)}x_i.$ This action is denoted by $(\bB, \bve)\cdot g$ and it replaces each variable $x_i$ of $g$ by a "new"variable $y_i$
\[
y_i = x_i + \sum_{j=0}^{i-1} b_{i,j} x_j + \varepsilon_i.
\]
Note that $y_i$ is a \textit{linear form}, i.e., a polynomial of degree 1. Also, the maximum variable of this linear form is $x_{i}$, as others are smaller than $x_i$ w.r.t. the order relation $\preceq.$
{\color{black}
Moving forward, we can recall the notion of orbits.\begin{definition}[\cite{bardet,dragoi17thesis}]
    The orbit of a monomial $f$ under the action of $\Alow$ is 
    defined as the set of polynomials 
    \[
        \Alow \cdot f = \{(\bB,\bve) \cdot f\mid (\bB,\bve) \in \Alow\}.
    \]
\end{definition}

Since $\Alow$ acts as a permutation on $\ev(f)$, all the elements in $\Alow\cdot f$ have the same Hamming weight. 

\subsection{Minimum Weight Codewords}\label{ssec:wmin_orbit}
In \cite{bardet} the authors demonstrated that any decreasing monomial code $\C(\I)$ with $r=\max_{f\in \I}\deg(f)$ has minimum distance $\wm\triangleq 2^{m-r}$ and any minimum weight codeword of $\C(\I)$ lie in one of the orbits $\Alow\cdot f$ where $f\in\I_r.$ The main challenge was to determine the stabilizer subgroup for each coset leader. Then, a particular subgroup of $\Alow$ came into play, subgroup that will be useful for our case as well.

\begin{definition}[\cite{bardet,dragoi17thesis}] For any $g\in\Mon$ define $\Alow_g$ as the subgroup $\Alow$ where 
\[
 \varepsilon_i = 0 \text{ if } i \in \ind(\check{g}) 
 ~~~\text{ and }~~ 
 b_{ij} = 
 \left \{ 
 	\begin{array}{l}
		0 \text{ if } i \in  \ind(\check{g}) \\
		0 \text{ if }  j \in \ind(g).
	 \end{array}
 \right.
\] 
\end{definition}

{\color{black}

This subgroup spans the whole orbit as provided by the following theorem.
\begin{theorem}[\cite{bardet,dragoi17thesis}]\label{thm:min-weight}
Let $f\in \Mon.$ Then we have \begin{equation}
\Alow\cdot f=\Alow_f\cdot f.
\end{equation}

Moreover, there are no polynomials in $\Alow_f\cdot f$ that are fixed by more than one group element (the identity).
\end{theorem}

Estimating the size of an orbit boils down to counting the number of $\varepsilon_i$ (translations) and $b_{i,j}$ (linear mapping) from $\Alow_f.$ For translations we have $2^{\deg(f)}$ choices, while for linear mapping we have $2^{|\lambda_f|}$ choices \cite{bardet}. To be more precise, for $f=x_{i_{1}}\dots x_{i_{r}}$ with $i_1<\dots< i_r$ the number of free variables taken together with $x_i\in\ind(f)$ is $\lambda_f(x_i)=|J_f(i)|,$ where $J_f(i)=\{j\in[0,i) \mid j\notin\ind(f)\}.$ One can associate a partition to $f$, of length $\deg(f)$ defined by $\lambda_f=(i_{r}-(r-1),\dots,i_1-0).$ Hence, the number of free variables on all $x_i$ in the set $\ind(f)$ equals $|\lambda_f(f)|=\sum_{i\in\ind(f)}\lambda_f(x_i)$, (or simply $\lambda_f$).

    In \cite{vlad1.5d} the definition of $\lambda_f$ was extended to any monomial $g=x_{j_{0}}\dots x_{j_{l-1}}$ satisfying $g|f$, i.e., $\lambda_f(g)$ is the partition of length $l$ defined by $\lambda_f(g)=(\lambda_f(x_i))_{i\in\ind(g)}$, which yields $ |\lambda_f(g)|=\sum_{i\in\ind(g)}\lambda_f(x_i).$ In other words $\lambda_f(g)$ is the restriction of $\lambda_f(f)$ to the indices from $\ind(g).$ 


    
    

 Finally, we get the well known formula from \cite{bardet}
\begin{equation}\label{eq:A_wm}
    \left|\Alow_f \cdot f\right| = 2^{\deg(f)+|\lambda_f|}.
\end{equation}
The collection of all orbits $\Alow_f\cdot f$ for $f\in\I_r$ gives the complete characterization of minimum weight codewords of any decreasing monomial code \cite{bardet}.

In our article, any code $\C(\I)$ will implicitly be decreasing monomial, and the parameters $r$ always denote the maximum degree of monomials in $\I$, i.e., $r=\max_{f\in\I}\deg(f).$

\section{Codewords in terms of evaluation of Minkowski sums of orbits of maximum degree monomials}

Here we shall recall existing results on the classification, thus including structural properties, of higher than minimum weight codewords. We shall start with some fundamental results from the 70s related to Reed-Muller codes. We will reinterpret these results in terms of $\Alow$ and review existing results on higher weights codewords for decreasing monomial codes.

\subsection{Reed-Muller codes: classification results} 
\begin{theorem}
[{\cite{sloane1970weight},\cite[Theorem 1]{kasami1970weight}}]
\label{thm:Kasami-Tokura}
 Let $r<m$ and $P\in\Rm$ be such that $\deg(P)\leq r$ with $0<\w(\ev(P))<2^{m+1-r}.$ Then $P$ is affine equivalent to one of the forms 
\begin{enumerate}
	\item \textbf{Type I}: $P=y_1\dots y_{r-\mu}(y_{r-\mu+1}\dots y_{r}+y_{r+1}\dots y_{r+\mu})$ where $m\geq r+\mu,r\geq \mu\geq 3$
	\item \textbf{Type II}: $P=y_1\dots y_{r-2}(y_{r-1}y_{r}+\dots+y_{r+2\mu-3}y_{r+2\mu-2})$ where $m-r+2\geq 2\mu, 2\mu\geq 2.$
\end{enumerate} 

In both cases $y_i$ are linear independent forms and $\w(\ev(P))=\w_{\mu}\triangleq 2^{m+1-r}-2^{m+1-r-\mu}.$
\end{theorem}

The case $\mu=1$ boils down to $\wm$ since $r+2\mu-2=r.$ 

\begin{remark}
    Notice that $P$ can be rewritten as $P=P_1+P_2$ with $\w(\ev(P_1))=\w(\ev(P_2))=2^{m-r}$ since both $P_1$ and $P_2$ are minimum weight codewords. Indeed, $P_1=y_1\dots y_{r}$ and $P_2=y_1\dots y_{r-\mu}y_{r+1}\dots y_{r+\mu}$ are product of $r$ independent linear forms and thus fall inside the orbit of a minimum weight codeword.

    The weight of $P$ follows from the relation 
    \begin{align*}
     \w(\ev(P))&=\w(\ev(P_1+P_2))\\
     &=\w(\ev(P_1))+\w(\ev(P_2))-2\w(\ev(P_1P_2))\\
     &=2^{m-r}+2^{m-r}-2\w(\ev(P_1P_2))\\
     &=2^{m-r+1}-2^{1+m-r-\mu}=\w_{\mu}
     \end{align*}
     pointing out to $P_1P_2$ being a product of $r+\mu$ independent linear forms.  
\end{remark} 

A useful fact from Theorem \ref{thm:Kasami-Tokura} is that any codeword of weight $\w_{\mu}$ 
is defined by either $r+\mu$ independent linear forms (if Type I) or $r+2\mu-2$ independent linear forms (if Type II). 

Furthermore, given $\wm=2^{m-r}$ which was determined in \cite{bardet}, the resulting weight will be $\w_{\mu}=2^{m+1-r}-2^{m+1-r-\mu}=(2-1/2^{\mu-1})\wm$. Observe that $1.5\leq(2-1/2^{\mu-1})<2$ for $\mu\geq 2$.
The table below tabulates the available weights less than $2\wm$ for example codes and their corresponding types:
\begin{center}
\setlength{\tabcolsep}{2pt}
\begin{tabular}{ c|ccccc } 
 $\mu$ & {\color{blue}1} & 2 & 3 & 4\\
 \hline
 $\w_{\mu}$ & {\color{blue}$\wm$} & $1.5\wm$ & $1.75\wm$ & $1.875\wm$\\
 \hline
 For $\R(2,8)\!:\!(256,37)$ & {\color{blue}64} & 96 & 112 & 120\\
 Type & {\color{blue}II} & II & II &  II\\
 \hline
 For $\R(3,7)\!:\!(128,64)$ & {\color{blue}16} & 24 & 28 & -\\
 Type & {\color{blue}II} & II & I, II & -\\
\end{tabular}
\end{center}
Note that according to Theorem \ref{thm:Kasami-Tokura}, for a given $\R(r,m)$ code, we have the maximum $\mu$ as 
\begin{equation*}
\mu \leq \begin{cases}
\min(m-r,r) &\text{Type I}\\
\frac{m-r+2}{2} &\text{Type II}
\end{cases}
\end{equation*}
and the minimum as
\begin{equation*}
\mu \geq \begin{cases}
3 &\text{Type I}\\
1 &\text{Type II}
\end{cases}
\end{equation*}
Therefore, we cannot have $\mu>3$ for $\R(3,7)$ (corresponding to weight 30) and $\mu>4$ for $\R(2,8)$ (Type I does not exist as the upper bound $\mu\leq2$ is smaller than lower bound $\mu\geq3$). 

\begin{remark}
    Notice that all the results presented here are for codes satisfying $\R(1,m)\subset \CI.$ In other words, all decreasing monomials codes considered in this article satisfy $x_{m-1}\in\I.$    
\end{remark}

\begin{example}
    Let $m=9$ and $r=3.$ From the conditions in Thm. \ref{thm:Kasami-Tokura} we notice that $3\leq \mu\leq 3$ (Type I)) and $2\leq 2\mu\leq m-r+2=8$ which implies $\mu\leq 4$ (Type II)). Hence, we have 
    \begin{itemize}
        \item $\mu=1$ $P=y_1y_2y_3$ which gives $|\ev(P)|=\wm=2^{9-3}=64$;
        \item $\mu=2$ (Type II) in Thm. \ref{thm:Kasami-Tokura}), $P=y_1(y_2y_3+y_4y_5)$ and we have $|\ev(P)|=2^{7}-2^{7-2}=128-32=96$;
        \item $\mu=3$ (Type I) in Thm. \ref{thm:Kasami-Tokura}), $P=y_1y_2y_3+y_4y_5y_6$ and we have $|\ev(P)|=2^{7}-2^{7-3}=128-16=112$;
        \item $\mu=4$ (Type II) in Thm. \ref{thm:Kasami-Tokura}), $P=y_1(y_2y_3+y_4y_5+y_6y_7)$ and we have $|\ev(P)|=2^{7}-2^{7-4}=128-8=120$; 
    \end{itemize}
\end{example}

\subsection{Codewords in terms of $\Alow$ group action}
While describing codewords based on linear independence is useful for Reed-Muller codes, due to their permutation group, in the general case of decreasing monomial codes another description comes at hand. It was given in \cite{rowshan2024weight} (Theorem 2) and restates Kasami's theorem in terms of $\Alow.$ 
\begin{theorem}[\cite{rowshan2024weight}]\label{thm:Kasami-Tokura-1}
    Let $\CI$ be a decreasing monomial code with $r=\max_{f\in\I}\deg(f)$ and $P\in\mathrm{span}(I)$ be such that $0<\w(\ev(P))<2^{m+1-r}.$ Then 
\begin{enumerate}
	\item \text{Type I:} for $m\geq r+\mu,r\geq \mu\geq 3$\\
    $P=y_1\dots y_{r-\mu}(y_{r-\mu+1}\dots y_{r}+y_{r+1}\dots y_{r+\mu})$\\
    $P\in \Alow\cdot f+\Alow\cdot g$ with $f,g\in \Mon$ having $\deg(f)=\deg(g)=r.$  
	\item \text{Type II:} for $m-r+2\geq 2\mu\geq 2$\\
    $P=y_1\dots y_{r-2}(y_{r-1}y_{r}+\dots+y_{r+2\mu-3}y_{r+2\mu-2})$\\
    $P\in \sum_{i=1}^{\mu}\Alow\cdot f_i$ with $f_i\in\Mon$ satisfying $\deg(f_i)=r.$
\end{enumerate} 
Also, $y_i$ are linear independent forms and $\w(\ev(P))=\w_{\mu}.$
\end{theorem}

\text{Type II} codewords were characterized in \cite{rowshan2024weight} continuing on the same path as \cite{vlad1.5d} (where $\mu=2$ was determined). We shall recall this result for the sake of completeness. 

\begin{theorem}[\cite{rowshan2024weight}]  Let $\C(\I)$ be a decreasing monomial code and $r=\max_{f\in \I}\deg(f)$. Then, any \text{Type II} codeword of weight $\w_{\mu}$ 
with $m-r+2\geq 2\mu\geq 2$ belongs to 
        \[\Alow_h\cdot h\cdot\sum_{i=1}^{\mu}\Alow_{f_i} \cdot \frac{f_i}{h}\] 
        where $\forall i, f_i\in \I_r$, $h=\gcd(f_i,f_j)$, for all $i,j\in[1,\mu]$ ($i\neq j$) and $\deg(h)=r-2$.
\end{theorem}




\section{Characterization of {Type I} codewords}

We shall proceed step-by-step and firstly, have a deeper look at the existence conditions for Type I codewords. Our next result, details one of the implications, namely, if such a codeword exists then it has to belong to a particular Minkowski sum of orbits, that we characterize. This first part deals with the conditions on the monomials defining the orbits.

\subsection{Conditions on monomials}
Before formally stating our result, we first explain the key conditions. The whole idea behind it is that we require $r+\mu$ independent linear forms which means that we need at least $r+\mu$ distinct variables to obtain independence. Also, there are $r-\mu$ common factors between the two terms $\Alow\cdot f$ and $\Alow\cdot g.$ When $f$ and $g$ share $r-\mu$ common variable ($h=\gcd(f,g)$) things are rather straightforward, the remaining parts $f/h,g/h$ are co-prime and thus $\Alow$ acts trivially on both parts.   
If $f,g$ share more that $r-\mu$ common variables then we need to check which of those common variables are big enough to fetch using the action of $\Alow$ enough variables from $\widecheck{fg}.$ In other words, we can set a product of $r-\mu$ variables in $\gcd(f,g)$ (denoted $h$), and verify whether we can shift backwards the variables in $h^{*}=\gcd(f,g)/h$ (subtract a strictly positive integer from each index of the variables in $h$). If the shifted monomial divides $\widecheck{fg}$, in other words these variables are free to be considered in $\Alow$, then such combination is valid.\\
The last case to consider is when $f\notin \I.$ Then, the two monomials should be equal and we should eliminate all monomials smaller than $f$ that are not in $\I$ using the action of $\Alow$ on $f.$

\begin{theorem}\label{thm:equality-orbits}
    Let $\C(\I)$ be a decreasing monomial code and $r=\max_{f\in \I}\deg(f)$. Then any codeword $\ev(P)$ of Type I satisfying $\w(\ev(P))=\w_{\mu}$ 
    with  $m-r\geq \mu, r\geq \mu\geq 3$ is such that
    \begin{itemize}
        \item  
        \textbf{A.} $\exists f,g\in \I_r,h\in\I_{r-\mu}$ and $h\preceq_w\gcd(f,g)$ with
        \[P\in \Alow_h\cdot h\cdot\left(\Alow_f\cdot \frac{f}{h}+\Alow_g \cdot \frac{g}{h}\right),\] where 
        \begin{itemize}
        \item \textbf{A.1} either $h=\gcd(f,g)$  
        \item \textbf{A.2} or $hh^{*}=\gcd(f,g),h^{*}\neq 1$, and $\exists h_s^{*}\in \Mon$ satisfying $h_s^{*}\preceq_w \widecheck{fg}$ 
        and $h_s^{*}\prec_{sh} h^{*}$,
        \end{itemize} 
        \item \textbf{B.1} $\exists f\not\in \I_r$, $\deg(f)=r, h\in\I_{r-\mu},h\preceq_w f$ with
        \[P\in \Alow_h\cdot h\cdot\left( \Alow_f\cdot \frac{f}{h}+ \Alow_f \cdot \frac{f}{h}\right)\] 
        and the following conditions are satisfied 
        \begin{enumerate}
            \item for $i\in \ind(f), j \in \ind(\check{f})$ and $j<i$ it exists $x_jf/x_i\in \I_r$ and for all indices where $x_jf/x_i\not \in \I$ we have $(\bB,\varepsilon)\cdot f + (\bB^{*},\varepsilon^{*})\cdot f$ is such that $b_{i,j}=b_{i,j}^{*}$   
            \item $\exists h_s^{*}\in \Mon,h_s^{*}\preceq_w \check{f}$ 
            and $h_s^{*}\prec_{sh} f/h.$
            \end{enumerate}
    \end{itemize} 
\end{theorem}


\begin{example}
    Let $m=8$ and $r=3$ and $\mu=3.$
    \begin{itemize}
        \item \textbf{No shared variables}
        \begin{itemize}
            \item  $f=x_0x_1x_2$ and $g=x_3x_4x_5.$ We have $\gcd(f,g)=1$ and thus the only solution for $h$ is $h=1$ since $\deg(h)=r-\mu=0$, and thus $(f,g)$ is a valid pair.
        \end{itemize}
        \item \textbf{One shared variable} 
        \begin{itemize}
            \item $f=x_0x_1x_2$ and $g=x_0x_3x_4.$ We have $\gcd(f,g)=x_0$ and thus the only solution for $h$ is $h=1$ since $\deg(h)=r-\mu=0$, and thus $h^{*}=x_0.$ Finding $h_s^{*}\prec_{sh}h^{*}$ is impossible and thus $(f,g)$ is not a valid pair of monomials for type I codewords with $\mu=3.$
        \item $f=x_1x_2x_3$ and $g=x_1x_4x_5.$ We have $\gcd(f,g)=x_1$ and thus the only solution for $h$ is $h=1$ since $\deg(h)=r-\mu=0$, and thus $h^{*}=x_1.$ Let $h_s^{*}=x_0.$ We have $h_s^{*}\prec_{sh}h^{*}$ and $h_s^{*}\preceq_{w}\frac{x_0\dots x_7}{fg}=x_0x_6x_7.$ Thus $(f,g)$ is a valid pair for type I codewords.
        \end{itemize}
        \item \textbf{Two shared variables}
        \begin{itemize}
            \item $f=x_0x_2x_3$ and $g=x_0x_2x_4.$ We have $\gcd(f,g)=x_0x_2$ and $h=1$ since $\deg(h)=r-\mu=0$, and thus $h^{*}=x_0x_2.$ Finding $h_s^{*}\prec_{sh}h^{*}$ is impossible (due to $x_0$) and thus $(f,g)$ is not a valid pair of monomials for type I codewords with $\mu=3.$
            \item $f=x_1x_2x_3$ and $g=x_1x_2x_4.$ We have $\gcd(f,g)=x_1x_2$ and $h=1$ since $\deg(h)=r-\mu=0$, and thus $h^{*}=x_1x_2.$ Also, $\frac{x_0\dots x_7}{fg}=x_0x_5x_6x_7.$ It is impossible to find $h_s^{*}$ satisfying both $h_s^{*}\prec_{sh}x_1x_2$ and $h_s^{*}\preceq_w x_0x_5x_6x_7$, thus $(f,g)$ is not a valid pair of monomials for type I codewords with $\mu=3.$
            \item $f=x_1x_4x_5$ and $g=x_1x_4x_6.$ We have $\gcd(f,g)=x_1x_4$ and $h=1$ since $\deg(h)=r-\mu=0$, and thus $h^{*}=x_1x_4.$ Also, $\frac{x_0\dots x_7}{fg}=x_0x_2x_3x_7.$ Both $h_s^{*}=x_0x_2$ and $h_s^{*}=x_0x_3$ satisfy  $h_s^{*}\prec_{sh}x_1x_4$ and $h_s^{*}\preceq_w x_0x_2x_3x_7$. Hence $(f,g)$ is a valid pair of monomials for type I codewords with $\mu=3.$
        \end{itemize}
        \item \textbf{Three shared variables}
        \begin{itemize}
            \item $f=g=x_0x_2x_4.$ We have $\gcd(f,g)=x_0x_2x_4$ and $h=1$ since $\deg(h)=r-\mu=0$, and thus $h^{*}=f=x_0x_2x_4.$ Finding $h_s^{*}\prec_{sh}h^{*}$ is impossible (due to $x_0$) and thus $(f,f)$ is not a valid pair of monomials for type I codewords with $\mu=3.$
            \item $f=x_1x_3x_5$ and $g=x_1x_3x_5.$ We have $\gcd(f,g)=x_1x_3x_5$ and $h=1$ since $\deg(h)=r-\mu=0$, and thus $h^{*}=f.$ Also, $\frac{x_0\dots x_7}{fg}=x_0x_2x_4x_6.$ Thus $h_s^{*}=x_0x_2x_4$ satisfy  $h_s^{*}\prec_{sh}x_1x_3x_5$ and $h_s^{*}\preceq_w x_0x_2x_4x_6$. Hence $(f,f)$ is a valid pair of monomials for type I codewords with $\mu=3.$
        \end{itemize}
    \end{itemize}
\end{example}

     

While the aforementioned conditions on the monomials are necessary, they are not sufficient. Indeed, we have not yet included the restriction on the number of independent linear forms in the Minkowski sum. Take for example the case of a valid pair $f=x_1x_2x_3$ and $g=x_1x_4x_5$ with $m=8,r=3,\mu=3$ and non-trivial gcd. We need to take $h=1$ and thus a $\w_{\mu}$-weight codeword should lie inside the orbit $\Alow_f\cdot f+\Alow_g\cdot g.$ However, not all polynomials inside this sum are of weight $\w_{\mu}$, only those having $6$ independent linear forms. This particular orbit contains codewords with 4 different weights 
\begin{itemize}
    \item $1.5\wm$ or $\mu=2:$ $x_1(x_2x_3+x_4x_5)$ (Type II)  
    \item $1.75\wm$ or $\mu=3:$ $x_1x_2x_3+(x_1+x_0)x_4x_5$ (Type I)
    \item $2\wm:$ $x_1x_2x_3+(x_1+1)x_4x_5$
\end{itemize}

Adding the requirement on the number of independent linear forms will allow us to achieve our second step, the characterization of the orbits that generate $\w_{\mu}$-weight codewords. The challenge boils down to isolating the sub-orbits with this particular restriction.

\subsection{Orbit characterization}
In order to characterize the Minkowski sum or product of two orbits we need to define the concept of collisions.
\begin{definition}\label{def:collision}
    Collision: Let $P,P^{*}\in\Alow\cdot f$ and $Q,Q^{*}\in\Alow\cdot g.$ we say that 
    \begin{itemize}
        \item $(P,Q),(P^{*},Q^{*})$ produces a collision for addition if $P+Q=P^{*}+Q^{*}$ with $P\neq P^{*}, Q\neq Q^{*}.$
        \item $(P,Q),(P^{*},Q^{*})$ produces a collision for multiplication if $PQ=PQ^{*}$ or $PQ=P^{*}Q^{*}$ or $PQ=P^{*}Q$ with $P\neq P^{*}, Q\neq Q^{*}.$
    \end{itemize}
\end{definition}

\begin{example}
    Let $f=x_0x_1$ and $g=x_0x_2.$ A collision for addition is $x_0x_1+x_0x_2=x_0(x_1+1)+x_0(x_2+1).$ A collision for multiplication is $\left((x_0+1)x_1\right)\left(x_0(x_2+1)\right)=\left((x_0+1)(x_1+1)\right)\left(x_0x_2\right).$
\end{example}

\subsubsection{The sub-case \textbf{A.1}}
Recall that here we have $h=\gcd(f,g)$ with $\deg(h)=r-\mu$ and $\mu\geq 3.$ Notice that this is the simplest case since $\Alow_h\cdot h\times \Alow_f\cdot \frac{f}{h}\times \Alow_g\cdot \frac{g}{h}$ is a product of $r+\mu$ independent linear forms. Indeed, there are $r+\mu$ variables in $fg$ and thus for this case one simply needs to estimate the cardinal of the orbits in order to determine how many Type I codewords exist. Sub-case \textbf{A.1} is a particular case of the following general theorem.
\begin{theorem}\label{thm:Minkowski-sum}
    Let $\I$ be a decreasing monomial set and $f,g\in\I_r$, with $r=\max_{h\in\I}\deg(h).$ If any of the two following conditions holds
    \begin{enumerate}
        \item the first two maximum variables in the set $\ind\left(\frac{fg}{\gcd(f,g)}\right)$ belong to $\ind(f)$ or $\ind(g),$
        \item $\deg(\gcd(f,g))<r-2.$ 
    \end{enumerate}
    we have 
    \begin{multline}
        \left|\Alow\cdot f+\Alow\cdot g\right|=\\\left|\Alow\cdot f\right|\left|\Alow\cdot g\right|.
    \end{multline}
\end{theorem}

    Since sub-case \textbf{A.1} requires $h=\gcd(f,g)$ with $\deg(h)=r-\mu$ and $\mu\geq 3$, this implies $\deg(\gcd(f,g))<r-2$ and thus we satisfy the second condition from Theorem \ref{thm:Minkowski-sum}. We can now proceed to estimating the number of sub-case \textbf{A.1} Type I codewords.


\begin{proposition}\label{pr:count-orbits-typeI-sub1}
      Let $\C(\I)$ be a decreasing monomial code and $r=\max_{f\in \I}\deg(f)$. Let $m-r\geq \mu, r\geq \mu\geq 3$ and $f,g\in \I_r$ with $h=\gcd(f,g)$ and $\deg(h)=r-\mu.$ Then the number of $\w_{\mu}$-weight codewords of sub-type A.1 defined by $f,g$ is given by the size of the orbit, i.e.,
        \begin{multline}
            \left|\Alow_h\cdot h\cdot\left(\Alow_f\cdot \frac{f}{h}+\Alow_g \cdot \frac{g}{h}\right)\right| \\ = \left|\Alow_h\cdot h\right|\left|\Alow_f\cdot \frac{f}{h}\right|\left|\Alow_g \cdot \frac{g}{h}\right|. 
        \end{multline}
            
\end{proposition}


    Notice that while $\deg(\gcd(f,g))<r-2$ the condition on $h$ is crucial. Indeed, when $h\neq \gcd(f,g)$ the result from Proposition \ref{pr:count-orbits-typeI-sub1} can not be directly applied to estimate the size of the restricted orbit. The following example illustrates our claim.
    \begin{example}
        Let $m=8,r=4$ and $f=x_0x_1x_2x_3, g=x_0x_4x_5x_6.$ 
        \begin{itemize}
            \item for $\mu=3$ we have $\deg(h)=\deg(\gcd(f,g))=r-\mu=1.$ We have $h=x_0,f/h=x_1x_2x_3$ and $g/h=x_4x_5x_6$, while $f/h,g/h$ satisfy the condition from Proposition \ref{pr:count-orbits-typeI-sub1}. Hence, we have $r+\mu=7$ independent linear forms counted by $|\Alow\cdot x_0||\Alow_{f}\cdot x_1x_2x_3||\Alow_{g}\cdot x_4x_5x_6|.$
            \item for $\mu=4$ we have $\deg(h)=r-\mu=0$ while $\deg(\gcd(f,g))=1.$ Thus, $h=1,f/h=f$ and $g/h=g$ while $f,g$ satisfy the condition from Proposition \ref{pr:count-orbits-typeI-sub1}. However, there is an extra condition one needs to verify in this case. More exactly, the existence of $h_s^{*}\preceq_w \frac{x_0\dots x_7}{fg}=x_7$ and $h_s^{*}\prec_{sh} h^{*}=x_0$ which is impossible. In other words there are no polynomials in $\Alow\cdot f+\Alow\cdot g$ that satisfy the aforementioned condition. 
        \end{itemize} 
    \end{example}


\subsubsection{The sub-case \textbf{A.2}}

We shall informally explain what are the ingredients for this case. Suppose we have two monomials $f.g\in\I_r$ with $h\in\I_{r-\mu}$ such that $\gcd(f,g)=hh^{*}$ with $\deg(h^{*})\geq 1.$ The $\w_{\mu}$-weight codewords are evaluation of polynomials that are composed of $r+\mu$ independent linear forms lying inside the orbit $\Alow_h\cdot h\left(\Alow_f\cdot \frac{f}{h}+\Alow_g \cdot \frac{g}{h}\right).$ This means that we have to characterize exactly those polynomials that satisfying $\Alow_h\cdot h \times \Alow_f\cdot \frac{f}{h}\times \Alow_g \cdot \frac{g}{h}$ equals a product of $r+\mu$ independent linear forms. While $\Alow_h\cdot h \times \Alow_f\cdot \frac{f}{h}$ is a product of $r$ independent linear forms, the remaining part might not provide the extra $\mu$ independent linear forms, due to the existence of $h^{*}$ on both $f$and $g.$ Hence, the group action on $h^{*}$ should be restricted to only those matrices $\bB$ that produce independent linear forms with the preexisting forms coming from the action on $f.$ In other words, if on $f$ we allow the complete group action on the $h^{*}$, then on $g$ we need to split the action into two distinct parts, a free part on $\frac{g}{\gcd(f,g)}$ and a restricted part on $h^{*}.$  

Before we continue let us recall that for each variable $i\in\ind(f)$ the number of free variables from the group action is $J_f(i)=\{j\in[0,i) \mid j\notin\ind(f)\}.$ We thus have $(|J_{f}({i_l})|,\dots ,|J_{f}({i_1})|)=\lambda_{f}(h)$ for any divisor $h\preceq_w f$ with $\ind(h)=\{i_1,\dots ,i_l\}.$ Also, one can easily observe that the inclusion $J_{f}(i_1)\subseteq \dots J_{f}(i_l)$ holds.


\begin{definition}\label{def:restrict-LTA-inv}
    
    Let $f, g\in\Mon, h\preceq_w f$ and $h^{*}=\gcd(f/h,g)$ with $\deg(h^{*})=l$ and $\ind(h^{*})=\{i_1,\dots ,i_l\}.$ Let $(\bB,\varepsilon)\in\Alow_f$ be such that $\rank(\bB_{\ind(h^*),J_{fg}(i_{l})})=\deg(h^{*})$. The orbit $\Alow_f^{g}\cdot \frac{f}{h}$ is the collection of all polynomials of the form $(\bB,\varepsilon)\cdot \frac{f}{h}$ where $\bB$ satisfies the rank restrictions. 
    %
    
    
\end{definition}

This new group action can be explained as follows. Split the orbit $\Alow_f\cdot \frac{f}{h}$ into two separate actions, i.e., $\Alow_f\cdot \frac{f}{hh^{*}}$ and $\Alow_f\cdot h^{*}.$ This is always possible since the two monomials on which we act $\frac{f}{hh^{*}},h^{*}$ do not share any common variables. Because $h^{*}$ is the common factor with $g$ simply add a new restriction on this part of the group action. 

The orbit is well-defined even for the case $h=1.$

For the trivial case when $(f,g)$ or $(\frac{f}{h},g)$ are co-prime we have $\Alow_f^g\cdot \frac{f}{h}=\Alow_f \frac{f}{h}.$



{Let us give a formula for the cardinality of $\Alow_f^g\cdot f.$ }

\begin{proposition}\label{pr:coardinality-lta-gcd}
    Let $f,g\in\Mon,h\preceq_w f$ and $h^{*}=\gcd(f/h,g)$ with $\ind(h^{*})=\{i_1,\dots,i_l\}.$ Then as long as $\left|J_{fg}(i_j)\right|> j-1$ for $j\in\{1,\dots,l\}$ we have  
    \begin{equation}\label{eq:cardinal-lta-restirct-formula}
        \left|\Alow_f^g\cdot \frac{f}{h}\right|=\frac{2^{\deg(\frac{f}{h})+|\lambda_f\frac{f}{h}|}}{2^{|\lambda_{fg}h^{*}|}}\prod\limits_{j=1}^{\deg(h^{*})}\left(2^{|J_{fg}({i_j})|}-2^{j-1}\right).
    \end{equation}
\end{proposition}

Since we can split $|\lambda_f\frac{f}{h}|=|\lambda_f\frac{f}{hh^{*}}||\lambda_fh^{*}|$ we deduce that the first factor in \eqref{eq:cardinal-lta-restirct-formula} can be rewritten as
$2^{\deg(\frac{f}{h})+|\lambda_f\frac{f}{hh^{*}}| + |\lambda_fh^{*}|-|\lambda_{fg}h^{*}|}.$ Notice that 
\[|\lambda_fh^{*}|-|\lambda_{fg}h^{*}| = |\lambda_{\widecheck{\frac{g}{\gcd{(f,g)}}}}h^{*}|.\] 


\begin{remark}
    Notice that for any $h^{*}\neq 1$ we have
    \[\frac{\prod\limits_{j=1}^{\deg(h^{*})}\left(2^{|J_{fg}({i_j})|}-2^{j-1}\right)}{2^{|\lambda_{fg}h^{*}}|}=\prod\limits_{j=1}^{\deg(h^{*})}\left(1-\frac{1}{2^{|J_{fg}({i_j})|-(j-1)}}\right) < 1.\]
    Thus, naturally we get 
    \[\left|\Alow_f^g\cdot \frac{f}{h}\right|\leq 2^{\deg(f/h)+|\lambda_f\frac{f}{h}|}=\left|\Alow_f\cdot \frac{f}{h}\right|, 
    \]
    with equality when $\gcd(f/h,g)=1.$ 
\end{remark}

\begin{remark}
    When the group $\Alow_f$ acts on $f$ itself instead of $f/h$ we have
     \begin{equation}
        \left|\Alow_f^g\cdot {f}\right|=\frac{2^{\deg({f})+|\lambda_f{f}|}}{2^{|\lambda_{fg}\gcd{(f,g)}|}}\prod\limits_{j=1}^{\deg(\gcd{(f,g)})}\left(2^{|J_{fg}({i_j})|}-2^{j-1}\right)
    \end{equation}
    Also, if $g=f$ we have 
     \begin{equation}
        \left|\Alow_f^f\cdot \frac{f}{h}\right|=2^{\deg(\frac{f}{h})}\prod\limits_{j=1}^{\deg(\frac{f}{h})}\left(2^{|J_{f}({i_j})|}-2^{j-1}\right)
    \end{equation}

    with the particular case $h=1$
     \begin{equation}
        \left|\Alow_f^f\cdot {f}\right|=2^{\deg({f})}\prod\limits_{j=1}^{\deg(f)}\left(2^{|J_{f}({i_j})|}-2^{j-1}\right)
    \end{equation}
    
\end{remark}
In the example below we illustrate using Young diagrams how to compute the cardinality of the set $\Alow_f^g\cdot f.$  

\begin{example}\label{ex:young-diagrams}
Let $m=7$ and $f=x_1x_3x_5$ then the partition associated to $f$ is 
 $\lambda_f=(5-2,3-1,1-0)=(3,2,1)$ and it's Young diagram in the $3\times 3$ grid is 
    \begin{center}
    \begin{tikzpicture}[inner sep=0in,outer sep=5in]
      \node(m) {
\begin{ytableau}
 *(blue! 40) *  &  0 & 0 \\
 *(blue! 40) *  & *(blue! 40) * & 0 \\
*(blue! 40) *  &*(blue! 40) *   &*(blue! 40) *  \\
\end{ytableau}};
\node at (-0.5,1) {$x_0$};
\node at (0,1) {$x_2$};
\node at (0.5,1) {$x_4$};
\node at (-1,0.5) {$x_1$};
\node at (-1,0) {$x_3$};
\node at (-1,-0.5) {$x_5$};

\end{tikzpicture}.
 \end{center}
  where blue boxes represent free binary choices on the variables. We have $\left|\Alow_f\cdot f\right|=2^{3+6}=2^9.$
  
Let $g=f.$ Then $\gcd(f,g)=f$ with $J_f(1)=\{0\},J_{f}(3)=\{0,2\}, J_f(5)=\{0,2,4\}.$ We have the following diagram 
  \begin{center}
    \begin{tikzpicture}[inner sep=0in,outer sep=5in]
      \node(m) {
\begin{ytableau}
 *(red! 40) *  & 0 & 0 \\
 *(red! 40) *  & *(red! 40) * & 0 \\
*(red! 40) *  &*(red! 40) *   &*(red! 40) *  \\
\end{ytableau}};
\node at (-0.5,1) {$x_0$};
\node at (0,1) {$x_2$};
\node at (0.5,1) {$x_4$};
\node at (-1,0.5) {$x_1$};
\node at (-1,0) {$x_3$};
\node at (-1,-0.5) {$x_5$};
\end{tikzpicture}.
 \end{center}
   where, red boxes represent binary choices for linearly independent vectors. Hence, we have $\left|\Alow_f^f\cdot f\right|=2^{3}(2^1-2^{0})(2^2-2^1)(2^3-2^2)=2^6.$
   
Let $g=x_3x_5x_6$. Then $fg=x_1x_3x_5x_6$ and $h=\gcd(f,g)=x_3x_5$ with $J_{fg}(3)=\{0,2\}, J_{fg}(5)=\{0,2,4\}.$ We can compute $|\lambda_{fg}h|=|(3,2)|=5.$ The corresponding diagram is

    \begin{center}
    \begin{tikzpicture}[inner sep=0in,outer sep=5in]
      \node(m) {
\begin{ytableau}
 *(blue! 40) *  & 0 & 0 \\
 *(red! 40) *  & *(red! 40) * & 0 \\
*(red! 40) *  &*(red! 40) *   &*(red! 40) *  \\
\end{ytableau}};
\node at (-0.5,1) {$x_0$};
\node at (0,1) {$x_2$};
\node at (0.5,1) {$x_4$};
\node at (-1,0.5) {$x_1$};
\node at (-1,0) {$x_3$};
\node at (-1,-0.5) {$x_5$};
\end{tikzpicture}.
 \end{center}
   where, red boxes represent binary choices for linearly independent vectors. For example, at the position row $x_3$, column $x_2$ we cannot have $0.$ Hence, we have $\left|\Alow_f^g\cdot f\right|=2^{3+1}(2^2-2^{0})(2^3-2^1)=2^53^2.$

Changing the place of $f$ and $g$ leads to $\lambda_g=(6-2,5-1,3-0)=(4,4,3)$, and the following diagrams.
    \begin{center}
    \begin{tikzpicture}[inner sep=0in,outer sep=5in]
      \node(m) at (0.5,0) {
\begin{ytableau}
 *(blue! 40) *  &*(blue! 40) *  & *(blue! 40) *  & 0  \\
 *(blue! 40) *  &*(blue! 40) *  & *(blue! 40) * & *(blue! 40) *   \\
*(blue! 40) *  &*(blue! 40) *  &*(blue! 40) *   &*(blue! 40) *  \\
\end{ytableau}};
\node at (-0.25,1) {$x_0$};
\node at (0.25,1) {$x_1$};
\node at (0.75,1) {$x_2$};
\node at (1.25,1) {$x_4$};
\node at (-1,0.5) {$x_3$};
\node at (-1,0) {$x_5$};
\node at (-1,-0.5) {$x_6$};

\node(n) at (4.5,0){
\begin{ytableau}
 *(red! 40) *  &*(blue! 40) *  &*(red! 40) *   & 0 \\
 *(red! 40) *  & *(blue! 40) * & *(red! 40) *  &*(red! 40) *\\
*(blue! 40) *  &*(blue! 40) *   &*(blue! 40) * &*(blue! 40) *   \\
\end{ytableau}};
\node at (3.75,1) {$x_0$};
\node at (4.25,1) {$x_1$};
\node at (4.75,1) {$x_2$};
\node at (5.25,1) {$x_4$};
\node at (3,0.5) {$x_3$};
\node at (3,0) {$x_5$};
\node at (3,-0.5) {$x_6$};
\end{tikzpicture}.
\end{center}
This yields $\left|\Alow_g\cdot g\right|=2^{3+11}=2^{14}$ and $\left|\Alow_g^f\cdot g\right|=2^{3+6}(2^2-1)(2^3-2)=2^93^2.$
\end{example}

    \begin{proposition}\label{pr:count-orbits-typeI-sub2}
      Let $\C(\I)$ be a decreasing monomial code and $r=\max_{f\in \I}\deg(f)$. Let $m-r\geq \mu, r\geq \mu\geq 3$ and $f,g\in \I_r$ with $h|\gcd(f,g)$ with $\gcd(f,g)=hh^{*},\deg(h)=r-\mu$, such that $\exists h_s^{*}\in \Mon$ satisfying $h_s^{*}\preceq_w \widecheck{fg}$ 
      and $h_s^{*}\prec_{sh} h^{*}.$ Then, the number of $\w_{\mu}$-weight codewords of Type I sub-type A.2 defined by $f,g$ equals
         \[   \left|\Alow_h\cdot h\right|\left|\Alow_f\cdot \frac{f}{h}\right|\left|\Alow_g^{{f}{}} \cdot \frac{g}{h}\right|.
        \]
           When $f=g$ divide the formula by two.
\end{proposition}
\begin{remark}
    Notice that we have
    \begin{multline*}
        \left|\Alow_f\cdot \frac{f}{h}\right|\left|\Alow_g^{{f}{}} \cdot \frac{g}{h}\right|\\
        =\left|\Alow_g\cdot \frac{g}{h}\right|\left|\Alow_f^{{g}{}} \cdot \frac{f}{h}\right|.
    \end{multline*}    
\end{remark}

\begin{example}
    Let $m=7,r=3,\mu=3.$
    \begin{itemize}
        \item $f=g=x_1x_3x_5$ then we have a total of $2^{3+|\lambda_f|+3}(2^1-1)(2^2-2)(2^3-2^2)2^{-1}=16384$ polynomials $P$ with $\ev(P)=1.75\wm$ lying in the Minkowski sum $\Alow\cdot f+\Alow\cdot g.$
        \item $f=x_1x_3x_5, g=x_2x_4x_5$ then we have a total of $2^{3+|\lambda_f|+3+|\lambda_g|-|\lambda_{fg}x_5|}(2^1-1)=2^{19}$ codewords of weight $1.75\wm$
        \item $f=x_2x_3x_5,g=x_2x_3x_4$ then we have a total of $2^{3+|\lambda_f|+3+|\lambda_g|-|\lambda_{fg}(x_2x_3)|}(2^2-1)(2^2-2)=3\cdot2^{16}$ codewords of weight $1.75\wm$
    \end{itemize}
\end{example}
\subsubsection{The sub-case \textbf{B.1}}

Recall that here we need $f\not\in \I_r$, $\deg(f)=r, h\in\I_{r-\mu},h\preceq_w f$ 
        and the following holds 
        \begin{enumerate}
            \item for $i\in \ind(f), j\not \in \ind(f)$ and $j<i$ it exists $x_jf/x_i\in \I_r$ and for all indices where $x_jf/x_i\not \in \I$ we have $(\bB,\varepsilon)\cdot f + (\bB^{*},\varepsilon^{*})\cdot f$ is such that $b_{i,j}=b_{i,j}^{*}$   
            \item $\exists h_s^{*}\in \Mon,h_s^{*}\preceq_w \check{f}$ and $h_s^{*}\prec_{sh} f/h.$
            \end{enumerate}

We deal here with a case similar to sub-case \textbf{A.2}. We have two actions of $\Alow_f$ on $f$, one with no restrictions and one with several restrictions. The restrictions are \begin{itemize}
    \item with respect to monomials outside $\I:$ where $b_{i,j}$ are fixed, for all $x_jf/x_i\not\in\I$ where $i\in\ind(f), j<i, j\in\ind(\check{f}).$ 
    \item with respect to the number of independent linear forms: $\rank(\bB_{f,J_f(i_r)})=r.$
\end{itemize}

\begin{example}
    Let $m=8,r=3,\mu=3$ and $f=x_3x_6x_7\not\in\I.$ We have $\check{f}=x_0x_1x_2x_4x_5$ which implies $J_f(3)=\{0,1,2\},J_f(6)=J_f(7)=\{0,1,2,4,5\}$
    In order to have $\rank(\bB_{f,J_f(7)})=3$ we need at least $x_0x_6x_7,x_1x_3x_6,x_2x_3x_6\in\I.$ Indeed, if for example $x_0x_6x_7\not\in\I$ this would imply that on the first row there is no freedom of choice, i.e., all entries are fixed at zero, which means that the rank of our matrix is at most 2. 
\end{example}

The second action defines an orbit similar to $\Alow_f^f\cdot f.$ In order to define this new orbit we need to introduce the restriction of $J_f(i)$ to $\I$ denoted $J_f^{\I}(i)$ for any $i \in \ind(f)$, as follows $J_f^{\I}(i)=\{j\in[0,i-1) \mid j\not\in\ind(f),x_jf/x_i\in\I\}.$ Obviously, when $f\in\I$ we have $J_f^{\I}(i)=J_f(i)$ for all $i\in\ind(f)$, due to the decreasing property of $\I.$

\begin{definition}\label{def:restrict-LTA-inv}
    Let $f\not\in\Mon, h\preceq_w f$ with $\deg(h)=r-\mu$ and $\ind(f/h)=\{i_1,\dots ,i_{\mu}\}.$ Let $(\bB,\varepsilon)\in\Alow_f$ be such that $\rank(\bB_{\ind(f),J_{f}^{\I}(i_{\mu})})=\mu$. The orbit $\Alow_f^{f,\I}\cdot \frac{f}{h}$ is the collection of all polynomials of the form $(\bB,\varepsilon)\cdot {f}$ where $\bB$ satisfies the rank restrictions. 
    \end{definition}


\begin{proposition}\label{pr:cardinal-alow-restrictI}Let $f\not\in\Mon, h\preceq_w f$ with $\deg(h)=r-\mu$ and $\ind(f/h)=\{i_1,\dots ,i_{\mu}\}.$ Then as long as $\left|J_{f}^{\I}(i_j)\right|> j-1$ for all $j\in\{1,\dots,\mu\}$ we have
  \begin{equation}
        \left|\Alow_f^{f,\I}\cdot \frac{f}{h}\right|=2^{\deg(\frac{f}{h})}\prod\limits_{j=1}^{\deg(\frac{f}{h})}\left(2^{|J_{f}^{\I}({i_j})|}-2^{j-1}\right)
    \end{equation}    
\end{proposition}

  \begin{proposition}\label{pr:count-orbits-typeI-sub3}
      Let $\C(\I)$ be a decreasing monomial code and $r=\max_{f\in \I}\deg(f)$. Let $m-r\geq \mu, r\geq \mu\geq 3$ and $f\in \Mon,\deg(f)=r$ such that $f\not\in\I_r.$ Let $h\in\I_{r-\mu}$ with $h|f, \ind(f/h)=\{i_1,\dots,i_{\mu}\}$ s.t. $\exists h_s^{*}\in \Mon$ satisfying $h_s^{*}\preceq_w \check{f}$ and $h_s^{*}\prec_{sh} \frac{f}{h}.$ Then, as long as $\left|J_{f}^{\I}(i_j)\right|\geq j-1$ for all $j\in\{1,\dots,\mu\}$, the number of $\w_{\mu}$-weight codewords of Type I sub-type B.1 equals 
         \[   \left|\Alow_h\cdot h\right|\left|\Alow_f\cdot \frac{f}{h}\right|\left|\Alow_f^{{f,\I}{}} \cdot \frac{f}{h}\right|.
        \]
\end{proposition}

\begin{example}
    Let $m=7, r=4, \mu=3$ with $f=x_0x_2x_4x_6.$ In this case $h=x_0$ is the single valid option. Indeed, $f/h=x_2x_4x_6$ admits $h^{*}=x_1x_3x_5\preceq_w \check{f}=x_1x_3x_5$ and $h^{*}\prec_{sh}f/h.$ Any of the other possible cases, e.g., $h=x_2$ will necessarily imply $x_0\preceq_w f/h$ which comes into contradiction with $h^{*}\prec_{sh}f/h$ since $x_0$ is the smallest variable.\\
    So, let $h=x_0$ and $f/h=x_2x_4x_6.$ This implies $J_f(2)=\{1\},J_{f}(4)=\{1,3\},J_{f}(6)=\{1,3,5\}.$ Hence, if $x_0x_2x_4x_5,x_0x_2x_3x_6,x_0x_1x_4x_6\in\I_4$ then $J_f^{\I}(i)=J_{f}(i)$ for all $i\in\{2,4,6\}$ and the condition on $|J_f^{\I}(i_j)|> j-1$ is satisfied. In this case the counting is straightforward
    \begin{itemize}
        \item $|\Alow_h\cdot h|=2^1$
        \item $|\Alow_f\cdot \frac{f}{h}|=2^{3+1+2+3}=2^{9}$
        \item $|\Alow_f^{f,\I}\cdot \frac{f}{h}|=2^{3}(2^1-1)(2^2-2)(2^3-2^2)=2^{6}.$
    \end{itemize}

    Let $m=6,r=\mu=3$ and $f=x_2x_4x_5\not\in\I.$ In this case $h=1$ and $J_f(2)=\{0,1\},J_f(4)=\{0,1,3\},J_f(5)=\{0,1,3\}.$ \\
   Let's consider three cases, illustrated by the following Young tableau

    \begin{center}
    \resizebox{!}{0.35\columnwidth}{
    \begin{tikzpicture}[inner sep=0in,outer sep=5in]
      \node(m) at (0.5,0) {
\begin{ytableau}
 *(red! 40) *  &*(red! 40) *    & 0  \\
 *(red! 40) *  &*(red! 40) *   & $X$   \\
*(red! 40) *  &*(red! 40) *     &*(red! 40) *  \\
\end{ytableau}};
\node at (-0.05,1) {$x_0$};
\node at (0.45,1) {$x_1$};
\node at (0.95,1) {$x_3$};
%
\node at (-0.5,0.5) {$x_2$};
\node at (-0.5,0) {$x_4$};
\node at (-0.5,-0.5) {$x_5$};
\node at (0.45, 1.5) {\textbf{a)}};
\node at (0.45,-1.25) {$x_2x_3x_5\not\in\I$};
\node(n) at (4.5,0){
\begin{ytableau}
 *(red! 40) *  &$X$& 0 \\
 *(red! 40) *  & *(red! 40) *   &$X$\\
*(red! 40) *  &*(red! 40) *   &*(red! 40) *   \\
\end{ytableau}};
\node at (3.95,1) {$x_0$};
\node at (4.45,1) {$x_1$};
\node at (4.95,1) {$x_3$};
\node at (3.5,0.5) {$x_2$};
\node at (3.5,0) {$x_4$};
\node at (3.5,-0.5) {$x_5$};
\node at (4.45,1.5){\textbf{b)}};

\node at (4.45,-1.25){$x_2x_3x_5\not\in\I$};
\node at (4.45,-1.75){$x_1x_4x_5\not\in\I$};

\node(n) at (8.5,0){
\begin{ytableau}
 *(red! 40) *  &$X$& 0 \\
 *(red! 40) *  & *(red! 40) *   &$X$\\
*(red! 40) *  &*(red! 40) *   &$X$   \\
\end{ytableau}};
\node at (3.95+4,1) {$x_0$};
\node at (4.45+4,1) {$x_1$};
\node at (4.95+4,1) {$x_3$};
\node at (3.5+4,0.5) {$x_2$};
\node at (3.5+4,0) {$x_4$};
\node at (3.5+4,-0.5) {$x_5$};
\node at (4.45+4,1.5){\textbf{c)}};
\node at (4.45+4,-1.25){$x_2x_3x_5\not\in\I$};
\node at (4.45+4,-1.75){$x_1x_4x_5\not\in\I$};
\node at (4.45+4,-2.25){$x_2x_3x_4\not\in\I$};
\end{tikzpicture}
}.
\end{center}
The $X$-sign in the Young tableau denotes boxes that are fixed since the corresponding monomials are outside the set $\I.$ Red boxes denote free variables that are taken such thatthe rank condition is satisfied. 

\begin{itemize}
    \item \textbf{a)} $x_2x_3x_5\not\in\I$ while $x_2x_3x_4,x_1x_4x_5\in\I.$ We have $J_f^{\I}(2)=\{0,1\},J_f^{\I}(4)=\{0,1\},J_f^{\I}(5)=\{0,1,3\}$ and thus we can apply the formula 
    \begin{itemize}
        \item $|\Alow_f\cdot f|=2^{3+2+3+3}=2^{14}$
        \item $|\Alow_f^{f,\I}\cdot f|=2^{3}(2^2-1)(2^2-2)(2^3-2^2)=3*2^{6}$
    \end{itemize}
    \item \textbf{b)} $x_2x_3x_5,x_1x_4x_5\not\in\I$ while $x_2x_3x_4,x_1x_2x_5,x_0x_4x_5\in\I.$ We have $J_f^{\I}(2)=\{0\},J_f^{\I}(4)=\{0,1\},J_f^{\I}(5)=\{0,1,3\}$ and thus we can apply the formula 
    \begin{itemize}
        \item $|\Alow_f\cdot f|=2^{3+2+3+3}=2^{14}$
        \item $|\Alow_f^{f,\I}\cdot f|=2^{3}(2^1-1)(2^2-2)(2^3-2^2)=2^{6}$
    \end{itemize}
    \item \textbf{c)} We have $J_f^{\I}(2)=\{0\},J_f^{\I}(4)=\{0,1\},J_f^{\I}(5)=\{0,1\}$ and thus we can not apply the formula, in this case we don't have any $\w_{\mu}$-weight codewords as such.
\end{itemize}
\end{example}
\subsection{Counting formulae}




\begin{table*}[]
    \centering
    \begin{tabular}{c|c|c|c}
    \toprule
        \multicolumn{2}{c|}{Type} & Conditions & Cardinality of orbit \\
        \midrule
        \multicolumn{2}{c|}{Type I} &  $1\leq i\leq \mu,\;f_i\in \I_r,\quad  h=\gcd(f_i,f_j) \in \I_{r-2}$ & $\dfrac{2^{r-2+2\mu+ |\lambda_h|+\sum\limits_{i=1}^{\mu}|\lambda_{f_i}(\frac{f_i}{h})|}}{2^{\sum\limits_{(f_i,f_j)}\alpha_{\frac{f_i}{h},\frac{f_j}{h}}}}$\\
        \midrule
        \multirow{10}{*}{Type II} & \multirow{2}{*}{Sub-type A.1} & $f,g\in \I_r,\; h=\gcd(f,g),\;\deg(h)=r-\mu$& $2^{r+\mu+|\lambda_h|+\left|\lambda_f\left(\frac{f}{h}\right)\right|+\left|\lambda_g\left(\frac{g}{h}\right)\right|}$\\
        &&&\\
        \cline{2-4}
        &\multirow{4}{*}{Sub-type A.2}&&\multirow{4}{*}{${2^{r+\mu-\idfg+|\lambda_h|+\left|\lambda_f\left(\frac{f}{h}\right)\right|+\left|\lambda_g\left(\frac{g}{h}\right)\right|}}\prod\limits_{j=1}^{\deg(h^{*})}\left(1-\frac{1}{2^{|J_{fg}{(i_j)}|-(j-1)}}\right)$}\\
         & & $f,g\in \I_r,\;  \gcd(f,g)=hh^{*},\; \deg(h)=r-\mu,$ &\\
         && $h_s^{*}\preceq_w \widecheck{fg},\;h_s^{*}\prec_{sh} h^{*}$ &\\
         &&&\\
         \cline{2-4}
         &\multirow{4}{*}{Sub-type B.1}&&\multirow{4}{*}{${2^{r+\mu-1+|\lambda_h|+\left|\lambda_f\left(\frac{f}{h}\right)\right|}}
            \prod\limits_{j=1}^{\deg(h^{*})}\left({2^{|J_f^{\I}{(i_j)}|}-2^{j-1}}\right)$}\\
         & & $f\not\in \I\,;  \deg(f)=r,\; f=hh^{*},\; \deg(h)=r-\mu, $& \\
         &&$h_s^{*}\preceq_w \check{f},\; h_s^{*}\prec_{sh} \frac{f}{h}$&\\
         &&&\\
         \bottomrule
    \end{tabular}
    \caption{Counting formulae for $\w_{\mu}$-weight codewords of decreasing monomial codes}
    \label{tab:formula-all}
\end{table*}

In order to use the formulae for the orbits previously determined we need another result, namely the fact that a codeword of weight $\w_{\mu}$ belongs to only one sub-case. In other words the orbits are disjoint, fact that holds both for orbits of the same sub-type and for orbits of different sub-types. 

\begin{proposition}\label{pr:TypeI-disjoint}
    Let $\C(\I)$ be a decreasing monomial code and $r=\max_{f\in \I}\deg(f)$. Let $m-r\geq \mu, r\geq \mu\geq 3.$ Then any $\w_{\mu}$-weight codeword of Type I, say $\ev(P)\in\C(\I)$ belongs to a single sub-case. In other words, the orbits are disjoint both intra-cases as well as inter-cases. 
\end{proposition}

With this result at hand we can now give a complete formula for counting $\w_{\mu}$-weight codewords of Type I for any decreasing monomial code. 

\begin{theorem}\label{thm:count-all-typeI}
 Let $\C(\I)$ be a decreasing monomial code and $r=\max_{f\in \I}\deg(f)$. Let $m-r\geq \mu, r\geq \mu\geq 3$ and $\w_{\mu}=2^{m+1-r}-2^{m+1-r-\mu}.$ The number of weight $\w_{\mu}$ codewords of Type I is given by the formulae in Table \ref{tab:formula-all}.
\end{theorem}

\section{Applications}

\subsection{The Reed--Muller case}
Although weight enumeration formulae for codewords with weight smaller than $2\wm$ are known, these are computed in terms of orbits under the action of the complete affine group. We do compute these numbers in terms of orbits under $\Alow$ in order to validate our formulae. 

Notice that for Reed-Muller codes Type I codewords only come under sub-case A.1 and A.2. Indeed, there are no sub-case B.1 since all degree $r$ monomials are inside the $\I$ of a Reed-Muller code. 

\begin{table}[!ht]
    \centering
    \resizebox{0.49\textwidth}{!}{
    \begin{tabular}{c|c c c c}
       $m$ & $W_{\wm}$ & $W_{1.5\wm}$ & $W_{1.75\wm}$ & $W_{1.875\wm}$ \\
       &&&&\\
       \toprule
       \multicolumn{5}{c}{$\R(3,m)$}\\
       \midrule
         $m=7$ &  94488 & 74078592 & 3128434688 & 0  \\
         $m=8$ &  777240& 2698577280 & 304296714240 & 0  \\
         $m=9$ &  6304280& 91931532672& 27817105940480& 29533455515648 \\
         $m=10$ & 50781720& 3033740578176& 2661436632391680&  30212724992507904 \\
       \midrule
       \multicolumn{5}{c}{$\R(4,m)$}\\
       \midrule
       $m=7$ & 188976& 148157184& 5805342720& 0\\
       $m=8$ & 3212592& 12593360640& 1518742159360& 1684323434496 \\
       $m=9$ & 52955952& 919315326720& 271767121346560& 860689275027456\\
       $m=10$ & 859903792& 62697305282304& 43538373627330560& 313636859446034432\\
         \bottomrule
    \end{tabular}
    }
    \caption{Weight distribution for Reed--Muller codes}
    \label{tab:weight-RM-3-4}
\end{table}

\subsection{Polar codes}

Firstly we will consider $m=6$ and rate $0.5$ and $0.55.$ Both codes are subcodes of $\R(3,6).$ The rate $0.5$ polar code has all max degree monomials satisfying $f\preceq x_1x_3x_4, f\preceq x_0x_2x_5.$ In this case we have Type II codewords and Type I subcase A.1,A.2 codewords. Using our formula we deduce
$W_{8}=920,W_{12}=25472, W_{14}=32768.$ 

The rate $0.55$ polar code has max degree monomials satisfying $f\preceq x_1x_3x_5, f\preceq x_2x_3x_5.$ In this case we do have all sub-cases for Type I since the monomial $x_2x_3x_5\not\in\I$, however it is a valid monomial for Type I sub-case B.1. Our formula gives $W_{8}=2456, W_{12}=142208, W_{14}=868352.$ (Example 10 \cite{Ye2024-weightdistrib}).   

\subsection{Improving the partial order on weight contribution}

Our characterization can be used in order to further refine the weight contribution partial order $\preceq_{\wm}$ from \cite{rowshan2024weight}, i.e.,$f\preceq_{\wm} g \Leftrightarrow |\lambda_f|\leq |\lambda_g|.$
This order relation induces antichains that were characterized for sub-codes of $\R(2,m)$. Let us give a more general result regarding this matter.
\begin{lemma}
    Let $\A_{l,r}\triangleq\{f\in\Mon\mid \deg(f)=r\;, \sum_{i\in\ind(f)}i=l\}.$ Then any $f\neq g\in\A_{l,r}$ are non-comparable w.r.t. $\preceq_{\wm}.$ 
\end{lemma}

The question we face here is how to decide which monomials to consider from $\A_{l,r}$ when designing a decreasing monomial code with a better performance in terms of weight distribution. Our main idea is to further consider Type I codewords of higher weights. To be more precise, a monomial $f\in\A_{l,r}$ could generate $\w_{\mu}$-weight codewords of sub-type A.2. However, not all monomials in $\A_{l,r}$ generate the same amount of $\w_{\mu}$-weight codewords. Take the simplest case when $r=\mu$, which leads to
\[{2^{r+\mu-1+\left|\lambda_f\right|}}
            \prod\limits_{j=1}^{\deg(f)}\left({2^{|J_f^{f}{(i_j)}|}-2^{j-1}}\right).\]
Since all $f\in\A_{l,r}$ will have the same $|\lambda_f|$ the only thing that changes is the second term, namely the product. 
\begin{example}
    Let $m=8,r=3,\mu=3.$ We have that \\$\A_{11,3}=\{x_0x_4x_7,x_0x_5x_6,x_1x_3x_7,x_1x_4x_6,x_2x_3x_6,x_2x_4x_5\}.$ Monomials $x_0x_4x_7,x_0x_5x_6$ provides $0$ codewords of weight $\w_{3}.$ Using our formula we compute the second term, $2^37$ for $x_1x_3x_7$ while for the remaining monomials it equals $2^33^2.$    

We go even further and compute the RMxPolar code $\C(\I)$ defined by $\R(2,m)\subset\C(\I)\subset(\R(3,m))$ and $g\preceq f$ for any $g\in\I_3$, for all 6 cases $f\in\A_{11,3}.$
\begin{table}[!h]
    \centering
\begin{tabular}{c|c |ccc}
$f$&$|\I_3|$& $|W_{\wm}|$ & $|W_{1.5\wm}|$ & $|W_{1.75\wm}|$\\
\midrule
$x_0x_4x_7$ & 18&7000 &1694336& 26664960 \\
 $x_0x_5x_6$ & 15&5208& 583296& 1777664\\
 $x_1x_3x_7$ & 24&9240& 1975680& 23224320\\
 $x_1x_4x_6$ & 23&9240& 1975680& 23224320\\
 $x_2x_3x_6$ & 22&7960& 1323392& 14622720\\
  $x_2x_4x_5$ & 19& 7960& 1323392& 14622720\\
\bottomrule
\end{tabular}
    \caption{Weight Distribution for six RMxpolar codes}
    \label{tab:tab-exempl3}
\end{table}
Notice that $\C(\I)$ defined by $x_0x_4x_7$ presents an interesting property, although $|W_{\wm}|$ is close that of codes with similar dimension, the numbers $|W_{1.5\wm}|,|W_{1.75\wm}|$ are significantly bigger.
\end{example}



\ifCLASSOPTIONcaptionsoff
  \newpage
\fi

\bibliographystyle{plain}
\bibliography{refs}

\begin{thebibliography}{10}

\bibitem{bardet2016crypt}
M.~Bardet, J.~Chaulet, V.~Dragoi, A.~Otmani, and J.-P. Tillich.
\newblock Cryptanalysis of the {McEliece} public key cryptosystem based on
  polar codes.
\newblock In {\em Post-Quantum Cryptography, vol. 9606}, pages 118--143. 2016.

\bibitem{bardet}
M.~Bardet, V.~Dragoi, A.~Otmani, and J.-P. Tillich.
\newblock Algebraic properties of polar codes from a new polynomial formalism.
\newblock In {\em 2016 IEEE Int. Symp. Inf. Theory (ISIT)}, pages 230--234,
  2016.

\bibitem{dragoi17thesis}
Vlad~Florin Dragoi.
\newblock {\em An algebraic approach for the resolution of algorithmic problems
  raised by cryptography and coding theory}.
\newblock PhD thesis, Normandie Universit{\'e}, 2017.

\bibitem{Dragoi-Szocs}
Vlad-Florin Dr{\u{a}}goi and Andreea Szocs.
\newblock Structural properties of self-dual monomial codes with application
  to code-based cryptography.
\newblock In Maura~B. Paterson, editor, {\em Cryptography and Coding}, pages
  16--41, Cham, 2021. Springer International Publishing.

\bibitem{vlad1.5d}
Vlad-Florin Drăgoi, Mohammad Rowshan, and Jinhong Yuan.
\newblock On the closed-form weight enumeration of polar codes: 1.5d -weight
  codewords.
\newblock {\em IEEE Transactions on Communications}, 72(10):5972--5987, 2024.

\bibitem{fossorier}
M.~P.~C. Fossorier and Shu Lin.
\newblock Weight distribution for closest coset decoding of |u|u+v| constructed
  codes.
\newblock {\em in IEEE Trans. on Information Theory}, 43(3):1028--1030, May
  1997.

\bibitem{GEECB21}
M.~Geiselhart, A.~Elkelesh, M.~Ebada, S.~Cammerer, and S.~t~Brink.
\newblock Automorphism ensemble decoding of {Reed-Muller} codes.
\newblock {\em IEEE Trans. Commun.}, 69(10):6424--6438, October 2021.

\bibitem{GEEB21}
M.~Geiselhart, A.~Elkelesh, M.~Ebada, S.~Cammerer, and S.~ten Brink.
\newblock On the automorphism group of polar codes.
\newblock pages 1230--1235. 2021 IEEE Int. Symp. Inf. Theory (ISIT), July 2021.

\bibitem{IU22}
K.~Ivanov and R.~L. Urbanke.
\newblock On the efficiency of polar-like decoding for symmetric codes.
\newblock {\em IEEE Transactions on Communications}, 70(1):163--170, January
  2022.

\bibitem{arikan}
E.~Ar\i kan.
\newblock Channel polarization: A method for constructing capacity-achieving
  codes for symmetric binary-input memoryless channels.
\newblock {\em IEEE Trans. Inf. Theory}, 55(7):3051--3073, July 2009.

\bibitem{arikan2}
E.~Ar\i kan.
\newblock From sequential decoding to channel polarization and back again.
\newblock preprint, 2019.

\bibitem{kasami1970weight}
T.~Kasami and N.~Tokura.
\newblock On the weight structure of {Reed-Muller} codes.
\newblock {\em Trans. Inf. Theory}, 16(6):752--759, November 1970.

\bibitem{kasami1976w2.5d}
T.~Kasami, N.~Tokura, and S.~Azumi.
\newblock On the weight enumeration of weights less than 2.5d of reed---muller
  codes.
\newblock {\em Information and Control}, 30(4):380--395, 1976.

\bibitem{li2021complete}
Yuan Li, Huazi Zhang, Rong Li, Jun Wang, Wen Tong, Guiying Yan, and Zhiming Ma.
\newblock The complete affine automorphism group of polar codes.
\newblock In {\em 2021 IEEE Global Communications Conference (GLOBECOM)}, pages
  01--06. IEEE, 2021.

\bibitem{lin_costello}
S.~Lin and D.~J. Costello.
\newblock Error control coding.
\newblock In {\em 2nd Edition, Pearson}, pages 395--400. Prentice Hall, Upper
  Saddle River, 2004.

\bibitem{liu_analys}
Z.~Liu, K.~Chen, K.~Niu, and Z.~He.
\newblock Distance spectrum analysis of polar codes.
\newblock {\em in IEEE Wireless Communications and Networking Conference
  (WCNC)}, 2014:490--495, 2014.

\bibitem{ma2024-permutations}
Jicheng Ma and Guiying Yan.
\newblock On automorphism groups of polar codes, 2024.

\bibitem{PBL21}
C.~Pillet, V.~Bioglio, and I.~Land.
\newblock Polar codes for automorphism ensemble decoding.
\newblock {\em IEEE Information Theory Workshop (ITW}, 2021:1--6, October 2021.

\bibitem{polyanskaya}
R.~Polyanskaya, M.~Davletshin, and N.~Polyanskii.
\newblock Weight distributions for successive cancellation decoding of polar
  codes.
\newblock {\em IEEE Trans. Commun.}, 68(12):7328--7336, December 2020.

\bibitem{rowshan-pac1}
M.~Rowshan, A.~Burg, and E.~Viterbo.
\newblock Polarization-adjusted convolutional ({PAC}) codes: Sequential
  decoding vs list decoding.
\newblock {\em in IEEE Trans. on Vehicular Technology}, 70(2):1434--1447,
  February 2021.

\bibitem{rowshan2023formation}
Mohammad Rowshan, Son~Hoang Dau, and Emanuele Viterbo.
\newblock On the formation of min-weight codewords of polar/pac codes and its
  applications.
\newblock {\em IEEE Trans. on Inf. Theory}, 69(12):7627--7649, 2023.

\bibitem{rowshan2024weight}
Mohammad Rowshan, Vlad–Florin Drăgoi, and Jinhong Yuan.
\newblock Weight structure of low/high-rate polar codes and its applications.
\newblock In {\em 2024 IEEE Int. Symp. on Inf. Theory (ISIT)}, pages
  2945--2950, 2024.

\bibitem{rowshan2024channel}
Mohammad Rowshan, Min Qiu, Yixuan Xie, Xinyi Gu, and Jinhong Yuan.
\newblock Channel coding toward 6g: Technical overview and outlook.
\newblock {\em IEEE Open Journal of the Communications Society}, 5:2585--2685,
  2024.

\bibitem{rowshan2023minimum}
Mohammad Rowshan and Jinhong Yuan.
\newblock On the minimum weight codewords of pac codes: The impact of
  pre-transformation.
\newblock {\em IEEE Journal on Selected Areas in Information Theory},
  4:487--498, 2023.

\bibitem{sloane1970weight}
N.~J. Sloane and E.~R. Berlekamp.
\newblock The weight enumerator for second-order {R}eed-{M}uller codes.
\newblock {\em IEEE Trans. Information Theory}, 16(6):745--751.

\bibitem{valipour}
M.~Valipour and S.~Yousefi.
\newblock On probabilistic weight distribution of polar codes.
\newblock {\em IEEE commun. lett.}, 17(11):2120--2123, 2013.

\bibitem{yao}
Hanwen Yao, Arman Fazeli, and Alexander Vardy.
\newblock A deterministic algorithm for computing the weight distribution of
  polar codes.
\newblock In {\em 2021 IEEE International Symposium on Information Theory
  (ISIT)}, pages 1218--1223, 2021.

\bibitem{Ye2024-weightdistrib}
Zicheng Ye, Yuan Li, Huazi Zhang, Jun Wang, Guiying Yan, and Zhiming Ma.
\newblock On the distribution of weights less than 2wminin polar codes.
\newblock {\em IEEE Transactions on Communications}, 72(10):5988--6000, 2024.

\bibitem{zhang_prob}
Q.~Zhang, A.~Liu, and X.~Pan.
\newblock An enhanced probabilistic computation method for the weight
  distribution of polar codes.
\newblock {\em IEEE Communications Letters}, 21(12):2562--2565, 2017.

\end{thebibliography}

\appendices

\section{Technical facts}
\begin{lemma}[Proposition 3.7.12~\cite{dragoi17thesis}]\label{lem:prod-lin-form-distinct-var}
    Let $P=\prod_{j=1}^{l}y_j$ be a product of $l$ independent linear forms $y_j$ each having maximum variables $x_{i_j}$ (with respect to $\preceq$). Then $P$ can be written as $P=\prod_{j=1}^{l}y_j^{*}$ where all maximum variables $x_{i_j^{*}}$ in $y_j$ are pairwise distinct.
\end{lemma}

Straightforward, notice that the total number of distinct variables in a product of $l$ independent linear forms should always be at least equal to $l.$ 

Also, the proof of Theorem \ref{thm:Kasami-Tokura-1} comes directly from Theorem \ref{thm:Kasami-Tokura} and Lemma \ref{lem:prod-lin-form-distinct-var}.

\begin{definition}[Restricted orbits]
    Let $f\in\Mon$ with $\ind(f)=\{i_1,\dots,i_s\}$ and $P\in \Alow\cdot f.$ Define the restriction of $P$ to the subset $S\subseteq\ind(f)$ as \begin{equation*}
        P_{|S}=\prod_{i\in \ind(f)\setminus S}\left(x_i+\sum_{j<i,j \not \in \ind(f)}b_{i,j}x_j+\varepsilon_i\right).
    \end{equation*}
\end{definition}

\begin{lemma}[\cite{rowshan2024weight}]\label{lem:decomp-ltam-gcd}Let $f,g\in\I_r$ and $h=\gcd(f,g)\in\Mon.$ Then 
\begin{multline*}
    \Alow\cdot h\cdot \left(\Alow\cdot \frac{f}{h}+\Alow\cdot \frac{g}{h}\right)\\=
    \Alow_h\cdot h\cdot \left(\Alow_{f}\cdot \frac{f}{h}+\Alow_g\cdot \frac{g}{h}\right).
\end{multline*}
\end{lemma}


\section{Proof Of Theorem \ref{thm:equality-orbits}}

\begin{IEEEproof}
 By Theorem \ref{thm:Kasami-Tokura}, any codeword $\ev(P)$ of Type I satisfying the required weight condition can be written as
     $P=y_1\dots y_{r-\mu}(y_{r-\mu+1}\dots y_{r}+y_{r+1}\dots y_{r+\mu})$ where $m\geq r+\mu,r\geq \mu\geq 3.$ Let $\ev(P)\in\C(\I)$ be such a codeword 
    where $y_i$ are linear independent forms. We can write each term in $P$ as $y_j=x_{i_j}+l_j$ where $l_j$ is a linear form, sum of variables smaller than $x_{i_j}.$ \\
    Applying Lemma \ref{lem:prod-lin-form-distinct-var} to the product $y_1\dots y_{r-\mu}$ implies that all variables in the new product $y_1^{*}\dots y_{r-\mu}^{*}$ are pairwise distinct. Denote these variables by $x_{i_1},\dots,x_{i_{r-\mu}}.$ If any of the maximum variables from the remaining forms $y_{r-\mu+1},\dots y_{r+\mu}$, say $y_{r-\mu+j}$ with maximum variable $x_j$ are in the set $\{i_1,\dots,i_{r-\mu}\}$, i.e., $ x_j=x_{i_l}$ ($l<r-\mu$) then apply Lemma \ref{lem:prod-lin-form-distinct-var} by keeping $y_{i_l}^{*}$ in its initial form and modifying $y_{r-\mu+j}.$ Notice this is always possible, as consequence of Lemma \ref{lem:prod-lin-form-distinct-var}, since we have at least $r+\mu$ distinct variables in $y_1\dots y_{r+\mu}.$ \\  
    After applying Lemma \ref{lem:prod-lin-form-distinct-var} to all the common variables  we separate the two terms in $P=P_1+P_2$ with $P_1=y_1^{*}\dots y_{r}^{*}$, $P_2=y_1^{*}\dots y_{r-\mu}^{*}y_{r+1}^{\prime}\dots y_{r+\mu}^{\prime}$ where the maximum variables in each of the products $y_1^{*},\dots,y_r^{*}$ and $y_1^{*},\dots, y_{r-\mu}^{*},y_{r-\mu+1}^{\prime},\dots,y_{r+\mu}^{\prime}$ are pairwise distinct.\\
    Notice that there might be common maximum variables between $y_{r-\mu+1}^{*},\dots,y_{r}^{*}$ and $y_{r+1}^{\prime},\dots,y_{r+\mu}^{\prime}.$ 
    Now, let the maximum monomial be $x_{i_1}\dots x_{i_{r-\mu}}x_{i_{r-\mu+1}^{*}}\dots x_{i_{r}^{*}}$ for $P_1$ and $x_{i_1}\dots x_{i_{r-\mu}}x_{i_{r+1}^{*}}\dots x_{i_{r+\mu}^{*}}$ for $P_2$ and denote  \begin{align*}
        h&=\prod_{j\in [1,{r-\mu}]}x_{i_j}\\
         f&=h \prod_{j\in \{i_{r-\mu+1}^{*},\dots, i_{r}^{*}\}}x_j\\
         g&=h \prod_{j\in \{i_{r+1}^{*},\dots, i_{r+\mu}^{*}\}}x_j.
         \end{align*}    
         
    We have $P\in \Alow\cdot h\cdot \left(\Alow\cdot \frac{f}{h}+\Alow\cdot \frac{g}{h}\right).$ Since $P\in\mathrm{span}(\I)$ this implies that either i) both $f$ and $g$ belong to the set $\I$, or ii) $f=g\not \in \I.$ 
     
    \begin{enumerate}
        \item  if $f,g\in \I_r$ we have several cases. If $\gcd(f/h,g/h)=1$ the proof is finished. If not we have $\gcd(f/h,g/h)=h^{*}\neq 1.$ Since $P$ is made of $r+\mu$ linear independent forms we can multiply all the linear forms to obtain
        \begin{align*}
            P_1P_2&=\prod_{i\in \ind(h)}(x_{i}+l_{i})\prod_{i \in \ind(f/(hh^*)) }(x_{i}+l_{i})\\
            &\cdot \prod_{i\in\ind(g/(hh^*))}(x_{i}+l_{i})\\
            &\cdot\prod_{i\in \ind(h^{*})}(x_{i}+l_{i})(x_{i}+l_{i}^{*}),
        \end{align*}
        where using Lemma \ref{lem:prod-lin-form-distinct-var} for the last term we have
        \[\prod_{i\in \ind(h^{*})}(x_{i}+l_{i})(x_{i}+l_{i}^{*})=\prod_{i\in \ind(h^{*})}(x_{i}+l_{i})(l_{i}+l_{i}^{*}+1),\]
        where each $l_{i}+l_{i}^{*}+1$ has maximum variable $x_{j_i}\prec x_i$ for all $i\in \ind{h^{*}}.$ Hence $h_s^{*}\triangleq \prod_{i\in\ind{h^{*}}}x_{j_i}\prec_{sh} h^{*}$ and $h_s^{*}\preceq_w \frac{x_0\dots x_{m-1}}{fg}.$  
        \item 
    In the second case if $f,g\not\in\I$ we have $f=g.$ This implies

\begin{multline*}
        \left((\bB,\varepsilon)\cdot f\right)+\left( (\bB^{*},\varepsilon^{*})\cdot f\right)\\
        = \prod\limits_{j\in[1,r]}(x_{i_j}+l_{i_j})
        +\prod\limits_{j\in[1,r]}(x_{i_j}+l_{i_j}^{*})\\
        = \sum\limits_{j\in[1,r]} \frac{f}{x_{i_j}}l_{i_j}^{\prime}
        +\dots  +\prod\limits_{j\in[1,r]}l_{i_j}\prod\limits_{j\in[1,r]}l_{i_j}^{*},
            \end{multline*}
where $l_{i_j}^{\prime}\triangleq l_{i_j}+l_{i_j}^*.$
    If $\sum\limits_{j\in[1,r]} \frac{x_{i_{r-\mu+1}}\dots x_r}{x_{i_j}}(l_{i_j}^{\prime}) \in\mathrm{span}(\I)$ then all the other terms belong to $\mathrm{span}(\I)$ by definition of the $\Alow.$ Hence, we only need to analyze the restrictions on $\bB,\bB^{*}$ for having the aforementioned property. Notice that $f\not \in \left((\bB,\varepsilon)\cdot f\right)+\left( (\bB^{*},\varepsilon^{*})\cdot f\right).$ Without loss of generality let $x_{j_1}x_{i_2}\dots x_{i_{r}}\not \in \I$ with $j_1\not \in \ind(f)$ and $j_1\leq i_1.$ This implies that $g=x_{j_1}f/x_{i_1}$ and thus $x_{j_1}\not \in l_{i_1}^{\prime}$, or equivalently $x_{j_1}$ is nor in $l_{i_1}^{*}$ and $l_{i_1}$, or is both in $l_{i_1}^{*}$ and $l_{i_1}$, which is equivalent to  $b_{i_1,j_1}=b_{i_1,j_1}^{*}.$ Applying this to all monomials $g$ ends the first condition. For the second condition recall that we need $r+\mu$ distinct variables in order to satisfy the weight condition. Choosing $h\preceq_w f$ with $\deg(h)=r-\mu.$ Similar to the previous case let $h^{*}=f/h$ and thus one need to have $h_s^{*}\preceq_w \frac{x_0\dots x_{m-1}}{f}$ with $h_s^{*}\prec_{sh} h^{*}.$   
    \end{enumerate}
 \end{IEEEproof}

\section{Proof of Theorem \ref{thm:Minkowski-sum}}
We shall first begin with a technical lemma.
\begin{lemma}\label{lem:colision-to-deivisor}
    Let $\I\subseteq\Mon$ be a decreasing monomial set and $f,g\in I_r$ with a non-trivial common factor $h=\gcd(f,g).$ Then if $P,P^{*}\in\Alow\cdot f, Q,Q^{*}\in\Alow\cdot g$ are s.t. $P\neq P^{*}, Q\neq Q^{*}$ and $P-P^{*}=Q-Q^{*}$ we have 
    \begin{equation*}
    P_{|\ind(\frac{f}{h})}-P_{|\ind(\frac{f}{h})}^{*} = Q_{|\ind(\frac{f}{h})}-Q_{|\ind(\frac{f}{h})}^{*}    
    \end{equation*}
\end{lemma}

\begin{IEEEproof}
By Lemma \ref{lem:decomp-ltam-gcd}, we have 
 \begin{align*}
     P-P^{*}&=Q-Q^{*}\\
     P_{|\ind(\frac{f}{h})}H_f-P_{|\ind(\frac{f}{h})}^{*}H_f^{*}&=Q_{|\ind(\frac{g}{h})}H_g-Q_{|\ind(\frac{g}{h})}^{*}H_g^{*}
 \end{align*}
     where $P_{|\ind(\frac{f}{h})}\in\Alow_f\cdot{\frac{f}{h}}, H_{f}\in \Alow_f\cdot h$ (the same holds for $P^{*}$) and $Q_{|\ind(\frac{g}{h})}\in\Alow_g\cdot{\frac{g}{h}}, H_{g}\in \Alow_g\cdot h$ (the same holds for $Q^{*}$). Also, by definition the monomial $\prod_{i\in \ind(h)}x_{i}$ only appears in $H_f,H_f^{*},H_g,H_{g}^{*}.$ Thus, by extracting the coefficients of this monomial we deduce the wanted result.  
\end{IEEEproof}

We will split the proof of Theorem \ref{thm:Minkowski-sum} into two parts, each one considering one of the conditions in the theorem's statement. 
\subsection{First condition}
\begin{theorem}\label{thm:Minkowski-sum-first-condition}
    Let $\I\subseteq\Mon$ be a decreasing monomial set and $f,g\in \I_r$ s.t. $h=\gcd(f,g).$ Also, let  $f^{*}=f/h$ and $g^{*}=g/h$ s.t. $\ind(f^{*})=\{i_1,\dots,i_s\},\ind(g^{*})=\{j_1,\dots,j_s\}$ with $i_s>i_{s-1}>j_s$ or $j_s>j_{s-1}>i_s.$ Then 
\[|\Alow\cdot f+\Alow\cdot g|=|\Alow\cdot f||\Alow\cdot g|.\] 
\end{theorem}

To demonstrate Theorem \ref{thm:Minkowski-sum-first-condition} we will require an intermediate result. 

\begin{proposition}\label{pr:first-two-max-1}
Let $\I\subseteq\Mon$ be a decreasing monomial set and $f,g\in\I$ with $\ind(f)=\{i_1,\dots,i_s\},\ind(g)=\{j_1,\dots,j_s\}$ with $i_s>i_{s-1}>j_s$ or $j_s>j_{s-1}>i_s.$ Then \[|\Alow\cdot f+\Alow\cdot g|=|\Alow\cdot f||\Alow\cdot g|.\] 
\end{proposition}

\begin{IEEEproof}
Suppose the condition $i_s>i_{s-1}>j_s$ is satisfied. The proof works the same for the other condition. By absurd, suppose that there are two pairs of polynomials $(P,Q), (P^{*},Q^{*})\in \Alow \cdot f\times \Alow\cdot g$ with $P\neq P^{*},Q\neq Q^{*}$ such that $P-P^{*}=Q-Q^{*}$ or equivalently, $P-P^*+Q-Q^*=0.$ Let $P=(\bB,\bve)\cdot f,P^{*}=(\bB^{*},\bve^{*})\cdot f$ and $Q=(\bA,\bvg)\cdot g,Q^{*}=(\bA^{*},\bvg^{*})\cdot g$

 By definition we have  
\begin{align*}
P=\prod\limits_{k=1}^{s}(x_{i_k}+\sum\limits_{l<i_k, l\not\in\ind{f}}b_{i_k,l}x_{l}+\varepsilon_{i_k}),\\
Q=\prod\limits_{k=1}^{s}(x_{j_k}+\sum\limits_{l<j_k,l\not\in\ind{g}}a_{j_k,l}x_{l}+\gamma_{j_k})\\
P^{*}=\prod\limits_{k=1}^{s}(x_{i_k}+\sum\limits_{l<i_k,l\not\in\ind{f}}b_{i_k,l}^{*}x_{l}+\varepsilon_{i_k}^{*}),\\ 
Q^{*}=\prod\limits_{k=1}^{s}(x_{j_k}+\sum\limits_{l<j_k,l\not\in\ind{g}}a_{j_k,l}^{*}x_{l}+\gamma_{j_k}^{*}).
\end{align*}

Expanding the products and extracting the coefficient of the maximum variable, which is $x_{i_s}$, we obtain 
\begin{multline*}
    x_{i_s}(\prod\limits_{k=1}^{s-1}(x_{i_k}+\sum\limits_{l<i_k,l\not\in\ind{f}}b_{i_k,l}x_{l}+\varepsilon_{i_k})\\-\prod\limits_{k=1}^{s-1}(x_{i_k}+\sum\limits_{l<i_k,l\not\in\ind{f}}b_{i_k,l}^{*}x_{l}+\varepsilon_{i_k}^{*}))=0
\end{multline*}

Since the equation has to be valid for any $x_{i_s}\in\{0,1\}$ we deduce
\begin{multline}\label{eq:13}
\prod\limits_{k=1}^{s-1}(x_{i_k}+\sum\limits_{l<i_k,l\not\in\ind{f}}b_{i_k,l}x_{l}+\varepsilon_{i_k})\\-\prod\limits_{k=1}^{s-1}(x_{i_k}+\sum\limits_{l<i_k,l\not\in\ind{f}}b_{i_k,l}^{*}x_{l}+\varepsilon_{i_k}^{*})=0\\
\end{multline}
or equivalently $P_{|i_s}-P_{|i_s}^{*}=0.$
Since $P_{|i_s},P_{|i_s}^{*}\in\Alow_f\cdot \frac{f}{x_{i_s}}$, and $\Alow_f\cdot f$ does not admit non-trivial stabilizers, this implies equation $b_{i_k,l}=b_{i_k,l}^{*}$ and $\varepsilon_{i_k}=\varepsilon_{i_k}^{*}$ for all values of $l$ and $k<s$. Let us denote the linear factor by $y_{i_s}=(x_{i_s}+\sum\limits_{l<i_s,l\not\in\ind{f}}b_{i_s,l}x_{l}+\varepsilon_{i_s})$, i.e., $P=y_{i_s}P_{|i_s}$ and $P-P^{*}=(y_{i_s}-y_{i_s}^{*})P_{|i_s}.$ Notice that the maximum variable in $P_{|i_s}$ is $x_{i_{s-1}}.$ Hence, our initial equation becomes
\begin{equation}\label{eq:14}
Q-Q^{*}+P_{|i_s}\left(\sum\limits_{l<i_s,l\not\in\ind{f}}(b_{i_s,l}-b_{i_s,l}^{*})x_{l}+\varepsilon_{i_s}-\varepsilon_{i_s}^{*}\right)=0.
\end{equation}

The maximum variable in equation \eqref{eq:14} can be $x_{i_{s-1}}$ (when given by $P_{|i_s}$) or any other variable $x_{l}$ with $l>i_{s-1}$ (when given by $y_{i_s}-y_{i_s}^{*}$ ). 
 If $x_{i_{s-1}}$ is the maximum variable, this means that $b_{i_s,l}-b_{i_s,l}^{*}=0$ for any value of $i_{s-1}<l<i_s.$ By isolating the terms containing $x_{i_{s-1}}$ we obtain
\begin{equation}
x_{i_{s-1}}\left(\sum\limits_{l<i_{s-1},l\not\in\ind{f}}(b_{i_s,l}-b_{i_s,l}^{*})x_{l}+\varepsilon_{i_s}-\varepsilon_{i_s}^{*}\right)=0.
\end{equation}
This implies that $\varepsilon_{i_s}-\varepsilon_{i_s}^{*}=0$ and $b_{i_s,l}-b_{i_s,l}^{*}=0$ for any value of $l<i_{s-1}$, which implies $y_{i_s}=y_{i_s}^{*}$ and hence $P=P^{*}$ which ends the proof.

 If $x_{l}$ with $l>i_{s-1}$ is the maximum variable the for any $l>i_{s-1}$ we have
\begin{equation}
P_{|i_s}\left((b_{i_s,l}-b_{i_s,l}^{*})x_{l}+\varepsilon_{i_s}-\varepsilon_{i_s}^{*}\right)=0.
\end{equation}
So,  $b_{i_s,l}-b_{i_s,l}^{*}=0,\varepsilon_{i_s}-\varepsilon_{i_s}^{*}=0$ for any index $l>i_{s-1}.$ This leads to

\begin{equation}\label{eq:15}
Q-Q^{*}+P_{|i_s}\left(\sum\limits_{l<i_{s-1},l\not\in\ind{f}}(b_{i_s,l}-b_{i_s,l}^{*})x_{l}\right)=0.
\end{equation}

The maximum variable in equation \eqref{eq:15} is $i_{s-1}$ and hence we should have that $b_{i_s,l}=b_{i_s,l}^{*}$ for all values of $l$. But this implies both $P=P^{*}$ and $Q=Q^{*}$ and hence, finish our demonstration.
\end{IEEEproof}

With this result at hand we can proceed to the demonstration of Theorem \ref{thm:Minkowski-sum-first-condition}. 

\begin{IEEEproof}
    Suppose by absurd that we do have a pair of polynomials $(P,Q), (P^{*},Q^{*})\in \Alow \cdot f\times \Alow\cdot g$ with $P\neq P^{*},Q\neq Q^{*}$ such that $P-P^{*}=Q-Q^{*}$ or equivalently, $P-P^*+Q-Q^*=0.$ Let $P=(\bB,\bve)\cdot f,P^{*}=(\bB^{*},\bve^{*})\cdot f$ and $Q=(\bA,\bvg)\cdot g,Q^{*}=(\bA^{*},\bvg^{*})\cdot g$. By Lemma \ref{lem:colision-to-deivisor} we have 
     \begin{equation*}
    P_{|\ind(\frac{f}{h})}-P_{|\ind(\frac{f}{h})}^{*} = Q_{|\ind(\frac{f}{h})}-Q_{|\ind(\frac{f}{h})}^{*}.   
    \end{equation*}
    Since $f/h$ and $g/h$ satisfy the hypothesis of Proposition \ref{pr:first-two-max-1} we deduce that 
     \begin{equation}
    P_{|\ind(\frac{f}{h})}=P_{|\ind(\frac{f}{h})}^{*} \quad;\quad Q_{|\ind(\frac{f}{h})}=Q_{|\ind(\frac{f}{h})}^{*}.   
    \end{equation}
    Moving up to the initial condition we obtain
    \begin{equation}
 P_{|\ind(\frac{f}{h})}(H_f-H_f^{*})=Q_{|\ind(\frac{g}{h})}(H_g-H_g^{*})
 \end{equation}
     where $H_{f},H_{f}^{*}\in \Alow_f\cdot h$ and $H_{g},H_g^{*}\in \Alow_g\cdot h$
     \begin{align*}
         H_f&=\prod\limits_{i\in \ind(h)}^{}(x_{i}+\sum\limits_{l<i,l\not\in\ind{f}}b_{i,l}^{}x_{l}+\varepsilon_{i}^{})\\
         H_f^{*}&=\prod\limits_{i\in \ind(h)}^{}(x_{i}+\sum\limits_{l<i,l\not\in\ind{f}}b_{i,l}^{*}x_{l}+\varepsilon_{i}^{*})\\
         H_g&=\prod\limits_{i\in \ind(h)}^{}(x_{i}+\sum\limits_{l<i,l\not\in\ind{g}}a_{i,l}^{}x_{l}+\gamma_{i}^{})\\
         H_g^{*}&=\prod\limits_{i\in \ind(h)}^{}(x_{i}+\sum\limits_{l<i,l\not\in\ind{g}}a_{i,l}^{*}x_{l}+\gamma_{i}^{*})
     \end{align*}

     Let $i_o\in \ind(h).$ Then, since the monomial $\prod_{i\in \ind(h)\setminus\{i_o\}}x_i$ only appears once in each term, extracting the coefficient of this monomial implies
        \begin{multline}\label{eq:17}
 P_{|\ind(\frac{f}{h})}(\sum\limits_{l<i_{o},l\not\in\ind{f}}(b_{i_o,l}-b_{i_o,l}^{*})x_{l}+\varepsilon_{i_o}-\varepsilon_{i_o}^{*})\\=Q_{|\ind(\frac{g}{h})}(\sum\limits_{l<i_{o},l\not\in\ind{f}}(a_{i_o,l}-a_{i_o,l}^{*})x_{l}+\gamma_{i_o}-\gamma_{i_o}^{*})
        \end{multline}
The maximum monomial in the left part of \eqref{eq:17} is $(b_{i_o,l}-b_{i_o,l}^{*})x_{l}\prod_{i\in \ind(\frac{f}{g})}x_i$ while in the right part of the equation is $(a_{i_o,l}-a_{i_o,l}^{*})x_{l}\prod_{i\in \ind(\frac{g}{g})}x_i.$ Since the condition in our theorem states that $i_s>i_{s-1}>j_s$ ($\ind(f/h)=\{i_s,\dots,i_1\},\ind(g/h)=\{j_s,\dots,j_1\}$) the equality in \eqref{eq:17} can hold only if $b_{i_o,l}=b_{i_o,l}^{*},a_{i_o,l}=a_{i_o,l}^{*}$ and $\varepsilon_{i_o}=\varepsilon_{i_o}^{*},\gamma_{i_o}=\gamma_{i_o}^{*}.$ The same argument applies to the rest of the variables $x_i,i\in\ind(h)$, which implies $H_f=H_f^{*}$ and thus concludes our proof.
        
\end{IEEEproof}
\subsection{Second condition}

\begin{proposition}\label{pr:degree-two-minkowski2}
     Let $\I\subseteq\Mon$ be a decreasing monomial set and $f,g\in I_r$ s.t. $h=\gcd(f,g)$ with $\deg(h)= r-3.$ Then 
\[|\Alow\cdot f+\Alow\cdot g|=|\Alow\cdot f||\Alow\cdot g|.\] 
\end{proposition}

\begin{IEEEproof}
     Suppose we have a pair of polynomials $(P,Q), (P^{*},Q^{*})\in \Alow \cdot f\times \Alow\cdot g$ with $P\neq P^{*},Q\neq Q^{*}$ such that $P-P^{*}=Q-Q^{*}$ or equivalently, $P-P^*+Q-Q^*=0.$ By Lemma \ref{lem:decomp-ltam-gcd} we have 
     \begin{equation}\label{eq:24}
    P_{|\ind(\frac{f}{h})}-P_{|\ind(\frac{f}{h})}^{*} = Q_{|\ind(\frac{g}{h})}-Q_{|\ind(\frac{g}{h})}^{*}.   
    \end{equation}
 Let us simplify the notations and put $P_{{|\ind(\frac{f}{h})}}=P(h)$ and the same for the remaining three polynomials in \eqref{eq:24}. Notice that $\deg(P(h))=\deg(P(h)^{*})=\deg(Q(h))=\deg(Q(h)^{*})$. Suppose $\ind(f/h)=\{i_3,i_2,i_1\}$ and $\ind(g/h)=\{j_3,j_2,j_1\}$ and the indices are in decreasing order $i_3>i_2>i_1, j_3>j_2>j_1.$ Also, we can assume $i_3>j_3.$ Extracting the coefficient of the maximum variable $x_{i_3}$ leads to $P(h)_{|i_3}=P(h)_{|i_3}^{*}.$ This implies 
 \begin{multline}
     P(h)_{|_{i_3}}(\sum\limits_{l<i_3,l\neq i_2,l\neq i_1}(b_{i_3,l}-b_{i_3,l}^{*})x_l+\varepsilon_{i_3}-\varepsilon_{i_3})\\=Q(h)-Q(h)^{*}
 \end{multline}

 Since the maximum variable in $Q(h)$ and $Q(h)^{*}$ is $x_{j_3}$, we have $b_{i_3,l}-b_{i_3,l}=0$ for all $l>i_3.$ Moving down with the variables, we have two cases: $b_{i_3,j_3}-b_{i_3,j_3}=0$ and $b_{i_3,j_3}-b_{i_3,j_3}=1$.

 {\textbf{The case $b_{i_3,j_3}-b_{i_3,j_3}=0$}} leads to $Q(h)_{|j_3}=Q(h)_{|j_3}^{*}.$
Thus, we have
\begin{multline}
     P(h)_{|_{i_3}}(\sum\limits_{l<j_3,l\neq i_2,l\neq i_1}(b_{i_3,l}-b_{i_3,l}^{*})x_l+\varepsilon_{i_3}-\varepsilon_{i_3})\\= Q(h)_{|_{j_3}}(\sum\limits_{l<j_3,l\neq j_2,l\neq j_1}(a_{j_3,l}-a_{j_3,l}^{*})x_l+\gamma_{j_3}-\gamma_{j_3}).
\end{multline}
We are almost done with this case. Notice that for any index $l>\max(i_2,j_2)$ we need to have $b_{i_3,l}-b_{i_3,l}^{*}=a_{j_3,l}-a_{j_3,l}^{*}=0.$ Moreover, if the next maximum variable is $x_{i_2}$ (it belongs to $P(h)_{i_3}$) and we need to have $a_{j_3,i_2}-a_{j_3,i_2}=1.$ But then we have a monomial, either $x_{i_2}x_{j_2}x_{j_1}$ or $x_{i_2}x_{j_2}x_{i_1}$ that has coefficient 1, which is impossible. Also, if instead of $i_2$ the maximum was $j_2$ the same conclusion can be deduces. In conclusion, in this case \eqref{eq:24} is true only if $P_{|\ind(\frac{f}{h})}=P_{|\ind(\frac{f}{h})}^{*}$ which ends the proof.

\textbf{The case $b_{i_3,j_3}-b_{i_3,j_3}=1$} leads to 
\begin{equation}
    P(h)_{|_{i_3}}= Q(h)_{|_{j_3}}-Q(h)_{|j_3}^{*}.
\end{equation}

The case $i_2>j_2$ is impossible since the monomial $x_{i_2}x_{i_1}$ can be present only in $P(h)_{|_{i_3}}$.
The case $j_2>i_2$ leads to $Q(h)_{|_{j_3,j_2}}-Q(h)_{|j_3,j_2}^{*}.$ This implies 
\begin{multline}
     (x_{i_2}+\sum\limits_{l<i_2,l\neq i_1}b_{i_2,l}x_l+\varepsilon_{i_2})(x_{i_1}+\sum\limits_{l<i_1}b_{i_1,l}x_l+\varepsilon_{i_1})\\=
        (\sum\limits_{l<j_2,l\neq j_1}(a_{j_2,l}-a_{j_2,l}^{*})x_l+\gamma_{j_2}-\gamma_{j_2,l}^{*})(x_{j_1}+\sum\limits_{l<j_1}a_{j_1,l}x_l+\gamma_{j_1})
\end{multline}
The last equation is impossible to be satisfied since $j_1\neq i_2$ and $j_1\neq i_1$, which end the proof.
\end{IEEEproof}

\section{Proof of Proposition \ref{pr:count-orbits-typeI-sub1}}

To demonstrate our result we will require a technical lemma.
  \begin{lemma}\label{lem:gener-lem3}
         Let $\I$ be a decreasing set with $f,g\in \I_r$ with $h=\gcd(f,g)$ and $\deg(h)\leq r-2.$ Then \begin{multline}
        \left|\Alow_h\cdot h\cdot\left(\Alow_{f} \cdot \frac{f}{h}+\Alow_{g}\cdot \frac{g}{h}\right)\right|=\\
        \left|\Alow_h\cdot h\right|\times \left|\Alow_{f} \cdot \frac{f}{h}+\Alow_{g}\cdot \frac{g}{h}\right|
    \end{multline}
    \end{lemma}

The condition on the degree of the $\gcd(f,g)$ is absolutely vital. Take for example $f=x_1x_3, g=x_2x_3$ with $r=2.$ The condition is violated, indeed, $\gcd(f,g)=x_3$ with degree $r-1.$ We have several collisions for multiplication here.
\[x_3(x_1+x_3)=(x_3+x_1+x_2+1)(x_1+x_2).\]

\begin{IEEEproof}
    Suppose there are polynomials $H,H^{*}\in\Alow_h\cdot h$ and $P,P^{*}\in \Alow_{f}\cdot \frac{f}{h}+\Alow_g\cdot \frac{g}{h}$ s.t. $HP=H^{*}P^{*}.$ Since $h$ is a product of variables that are not present in $P$ or $P^{*}$ extracting the coefficient of $h$ from $HP$ and $H^{*}P^{*}$ implies $P=P^{*}.$ This implies $P(H+H^{*})=0.$ By definition we can set $H=\prod_{i\in \ind(h)}(x_{i}+l_{i})$, $H^{*}=\prod_{i\in \ind(h)}(x_{i}+l_{i}^{*}).$  
Hence, we obtain 
\[P\left(\sum\limits_{i\in\ind(h)} \frac{h}{x_{i}}l_{i}^{\prime}
        +\dots  +\prod\limits_{i\in\ind(h)}l_{i}\prod\limits_{i\in\ind(h)}l_{i}^{*}\right)=0\]
where $l_{i}^{\prime}\triangleq l_{i}+l_{i}^*.$ Since all the monomials $h/x_{i}$ are unique when expanding the second term of the previous equation, we deduce 
\[P\frac{h}{x_{i}}(l_{i}+l_{i}^{*})=0,\quad \forall i \in \ind(h).\]
By definition of the $\Alow$ we have that none of the variables in $\ind(h/x_i)$ is present in $P$ or $l_{i}+l_{i}^{*}.$ Extracting the coefficients of $h/x_{i}$ we obtain 
\begin{equation}\label{eq:gener-lem-3}
    P(l_{i}+l_{i}^{*})=0,\quad \forall i \in \ind(h),
\end{equation}

 which is valid only if $l_{i}+l_{i}^{*}=0$ for all $i$. 
 
\end{IEEEproof}

\begin{remark}\label{rem:gener-lem3}
    The case $\deg(\gcd(f,g))=r-1$ can lead to a similar result as in Lemma \ref{lem:gener-lem3}, under the following condition. Let $f=x_{i_1}\dots x_{i_s}\dots x_{i_r}$ with $i_1<\dots i_s<\dots i_r$ and $h=x_{i_1}\dots x_{i_s}$ ($h$ contains the smallest variables in $\ind(f)$). Then we have 
    \begin{multline}
        \left|\Alow_h\cdot h\cdot\left(\Alow_{f} \cdot \frac{f}{h}+\Alow_{g}\cdot \frac{g}{h}\right)\right|=\\
        \left|\Alow_h\cdot h\right|\times \left|\Alow_{f} \cdot \frac{f}{h}+\Alow_{g}\cdot \frac{g}{h}\right|
    \end{multline}
\end{remark}
Regarding Remark \ref{rem:gener-lem3} notice that if $h$ contains the smallest variables from $\ind(f)$ then $l_i$ and $l_i^{*}$ do not contain any variables from $\frac{f}{h}$ and $\frac{g}{h}$, which is impossible, since it is in conflict with \eqref{eq:gener-lem-3}. This means that \eqref{eq:gener-lem-3} can hold as long as $l_i=l_i^{*}$ which leads to the same conclusion. 

The proof of Proposition \ref{pr:count-orbits-typeI-sub1} becomes now a formality. 
\begin{IEEEproof}
    By Theorem \ref{thm:Minkowski-sum} we have 
        \begin{multline*}
        \left|\Alow_f\cdot \frac{f}{h}+\Alow_g\cdot \frac{g}{h}\right|=\\\left|\Alow_f\cdot \frac{f}{h}\right|\left|\Alow_g\cdot \frac{g}{h}\right|.
    \end{multline*}
Combined with Lemma \ref{lem:gener-lem3} we obtain the wanted result.
    
\end{IEEEproof}

\section{Proof of Proposition \ref{pr:coardinality-lta-gcd}}
\begin{IEEEproof}
Let $h=x_{i_1}\dots x_{i_l}.$ By definition of the sets $J_{fg}(i)$ we have that $\lambda_{fg}(h)[i]=|J_{fg}(i)|$ hence we are left to demonstrate that the number of binary matrices $\bB^{*}\triangleq \bB_{\ind(h),J_{fg}({i_l})}$ of full row rank equals $\prod_{j=1}^l\left(2^{|J_{fg}({i_j})|}-2^{j-1}\right).$ This is a classic formula adapted to our case, which is of matrices with fixed decreasing entries on each row. We shall prove it for the sake of completeness. So, start with the first row of the matrix. So. for $\bB^{*}[1]$ we have $2^{|J_{fg}({i_1})|}$ possible choices on the first $|J_{fg}({i_1})|$ positions while the remaining positions $|J_{fg}({i_l})|-|J_{fg}({i_1})|$ are all set to zero, by definition of $\bB^{*}.$ We need to subtract the all zero vector, which makes $2^{|J_{fg}({i_1})|}-1$ possibilities for the first row. For the second row there are $2^{|J_{fg}({i_2})|}$ possible choices (first $|J_{fg}({i_2})|$ positions are free while the remaining are set to zero) minus the linear combinations of the previous rows. There are $2$ dependent vectors with the previous rows, the first row and the all zero vector. Moving forward, the same procedure is repeated, hence, insuring that all rows are linearly independent.       
\end{IEEEproof}

\section{Proof of Proposition \ref{pr:count-orbits-typeI-sub2}}

\begin{IEEEproof}
    First, we need to demonstrate an equivalent of Lemma \ref{lem:gener-lem3} for $h|\gcd(f,g).$ For that we do need to demonstrate one inclusion. Let us be more precise. Let $P=H(P_f+P_g)\in\Alow_h\cdot h\cdot \left(\Alow_{f} \cdot \frac{f}{h}+\Alow_{g}\cdot \frac{g}{h}\right)$ with $H\in \Alow_h\cdot h$, $P_f\in \Alow_{f} \cdot \frac{f}{h}$ and $P_g=\Alow_{g} \cdot \frac{g}{h}.$ $P$, also has to satisfy the following condition. $HP_fP_g$ is a product of $r+\mu$ linear independent forms. Hence, what we need to demonstrate is that $P\in \Alow_h\cdot h\cdot \left(\Alow_{f} \cdot \frac{f}{h}+\Alow_{g}^f\cdot \frac{g}{h}\right).$ 

   Since $hh^{*}=\gcd(f,g)$ we can write 
    $P\in \Alow_h\cdot h\cdot \left(\Alow_{f} \cdot h^{*}\frac{f}{\gcd(f,g)}+\Alow_{g}\cdot h^{*}\frac{g}{\gcd(f,g)}\right).$ Recall that $HP_fP_g$ is a product of $r+\mu$ linear independent forms. Since $H$ is a product of $r-\mu$ linear independent forms each one having maximum variables $x_i, i \in \ind(h)$ which satisfies $\ind(h)\cap \ind(\frac{f}{h}\cdot\frac{g}{h})=\emptyset$ we deduce that $P_fP_g$ has to be a product of $2\mu$ independent linear forms. Also, $P_f$ is a product of $\mu$ linear independent forms each one having maximum variables $x_i, i \in \ind(\frac{f}{h})=\ind(h^{*})\cup \ind(\frac{f}{\gcd(f,g)}).$ Let $h^{*}=x_{i_1}\dots x_{i_s}.$ In order for $P_fP_g$ to be a product of $2\mu$ linear independent forms the product $\prod\limits_{i \in \ind(h^{*})}(\sum\limits_{j<i, j\not \in \ind(g)}b_{i,j}^{*}x_j+\sum\limits_{j<i, j\not \in \ind(f)}b_{i,j}x_j+\varepsilon_i+\varepsilon_i^{*}+1)$ should be equal to a product of $|\ind(h^{*})|$ independent linear forms having maximum variables $x_i\not\in\ind(fg)$ for all $i\in \ind(h^{*}).$ By definition of $\Alow$ we have that $P_fP_g=$
    \bigskip
    
    \begin{strip}
   {\small
    \begin{align*}
        &=\prod\limits_{i \in \ind(h^{*})}(x_i+\sum\limits_{j<i, j\not \in \ind(f)}b_{i,j}x_j+\varepsilon_i)\times\prod\limits_{i \in \ind(\frac{f}{\gcd(f,g)})}(x_i+\sum\limits_{j<i, j\not \in \ind(f)}b_{i,j}x_j+\varepsilon_i)\\
        &\times \prod\limits_{i \in \ind(h^{*})}(x_i+\sum\limits_{j<i, j\not \in \ind(g)}b_{i,j}^{*}x_j+\varepsilon_i^{*})
        \times \prod\limits_{i \in \ind(\frac{g}{\gcd(f,g)})}(x_i+\sum\limits_{j<i, j\not \in \ind(g)}b_{i,j}^{*}x_j+\varepsilon_i^{*})\\
        &=\prod\limits_{i \in \ind(h^{*})}(x_i+\sum\limits_{j<i, j\not \in \ind(f)}b_{i,j}x_j+\varepsilon_i)\times\prod\limits_{i \in \ind(\frac{f}{\gcd(f,g)})}(x_i+\sum\limits_{j<i, j\not \in \ind(f)}b_{i,j}x_j+\varepsilon_i)\\
        &\times \prod\limits_{i \in \ind(\frac{g}{\gcd(f,g)})}(x_i+\sum\limits_{j<i, j\not \in \ind(g)}b_{i,j}^{*}x_j+\varepsilon_i^{*}) \times\prod\limits_{i \in \ind(h^{*})}(\sum\limits_{j<i, j\not \in \ind(g)}b_{i,j}^{*}x_j+\sum\limits_{j<i, j\not \in \ind(f)}b_{i,j}x_j+\varepsilon_i+\varepsilon_i^{*}+1)\\
        &=\prod\limits_{i \in \ind(\frac{f}{h})}(x_i+\sum\limits_{j<i, j\not \in \ind(f)}b_{i,j}x_j+\varepsilon_i)
        \times \prod\limits_{i \in \ind(\frac{g}{\gcd(f,g)})}(x_i+\sum\limits_{j<i, j\not \in \ind(g)}b_{i,j}^{*}x_j+\varepsilon_i^{*})\\
        &\times \prod\limits_{i \in \ind(h^{*})}(\sum\limits_{\substack{j<i, j\not \in \ind(g)\\j\in \ind(f)}}b_{i,j}^{*}x_j+\sum\limits_{\substack{j<i, j\not \in \ind(f)\\j \in \ind(g)}}b_{i,j}x_j+\sum\limits_{j<i,j\not\in\ind(fg)}(b_{i,j}+b_{i,j}^{*})x_j+\varepsilon_i^{\prime})\\
       &=\prod\limits_{i \in \ind(\frac{f}{h})}(x_i+\sum\limits_{j<i, j\not \in \ind(f)}b_{i,j}x_j+\varepsilon_i)
        \times \prod\limits_{i \in \ind(\frac{g}{\gcd(f,g)})}(x_i+\sum\limits_{j<i, j\not \in \ind(g)}b_{i,j}^{*}x_j+\varepsilon_i^{*})\\
        &\times \prod\limits_{i \in \ind(h^{*})}\left(\sum\limits_{\substack{j<i, j\not \in \ind(g)\\j\in \ind(f)}}b_{i,j}^{*}\left(\sum\limits_{l<j,l\not \in\ind(f)}b_{j,l}x_j+\varepsilon_j+1\right)+\sum\limits_{\substack{j<i, j\not \in \ind(f)\\j \in \ind(g)}}b_{i,j}\left(\sum\limits_{l<j,l\not \in\ind(g)}b_{j,l}^{*}x_j+\varepsilon_j^{*}+1\right)+\sum\limits_{j<i,j\not\in\ind(fg)}(b_{i,j}+b_{i,j}^{*})x_j+\varepsilon_i^{\prime}\right)\\
       &=\prod\limits_{i \in \ind(\frac{f}{h})}(x_i+\sum\limits_{j<i, j\not \in \ind(f)}b_{i,j}x_j+\varepsilon_i)
        \times \prod\limits_{i \in \ind(\frac{g}{\gcd(f,g)})}(x_i+\sum\limits_{j<i, j\not \in \ind(g)}b_{i,j}^{*}x_j+\varepsilon_i^{*})\\
        &\times \prod\limits_{i \in \ind(h^{*})}\left(\sum\limits_{j<i,j\not\in\ind(fg)}b_{i,j}^{\prime}x_j+\varepsilon_i^{\prime}\right). 
        \end{align*}
}
\end{strip}
   
  Since $P_f$ is already a product of $\mu$ independent linear forms, we can assume without loss of generality that the coefficients $b_{i,j}$ are fixed. We can thus take $b_{i,j}=0$, which leads to  $\prod\limits_{i \in \ind(h^{*})}\left(\sum\limits_{j<i,j\not\in\ind(fg)}b_{i,j}^{*}x_j+\varepsilon_i^{*}\right)$ has to be a product of $|\ind(h^{*})|$ independent linear forms, which is equivalent to $(\bB^{*},\varepsilon^{*})\in \Alow_g$ such that $\rank(\bB_{\ind(h^{*}),J(h^{*})}^{*})=\deg(h^{*}).$

Now let us demonstrate is that there are no collisions for multiplication. For that, let $(H,P_f,P_g)\neq(H^{*},P_f^{*},P_g^{*})$ such that $H(P_f+P_g)=H^{*}(P_f^{*}+P_g^{*})$ with with $H,H^{*}\in \Alow_h\cdot h$, $P_f,P_f^{*}\in \Alow_{f} \cdot \frac{f}{h}$ and $P_g,P_g^{*}=\Alow_{g} \cdot \frac{g}{h}.$ By definition of $\Alow$ the variables in $h$ are not present in $P_f+P_g$ and $P_f^{*}+P_g^{*}.$ Hence, extracting the coefficient of $h$ from the equality gives
\begin{equation*}
    P_f+P_g=P_f^{*}+P_g^{*}
\end{equation*}
which implies \[(H+H^{*})(P_f+P_g)=0.\]
Further we can set $H=\prod_{i\in \ind(h)}(x_{i}+l_{i})$, $H^{*}=\prod_{i\in \ind(h)}(x_{i}+l_{i}^{*}).$  
Hence, we obtain 
\[(P_f+P_g)\left(\sum\limits_{i\in\ind(h)} \frac{h}{x_{i}}l_{i}^{\prime}
        +\dots  +\prod\limits_{i\in\ind(h)}l_{i}\prod\limits_{i\in\ind(h)}l_{i}^{*}\right)=0\]
where $l_{i}^{\prime}\triangleq l_{i}+l_{i}^*.$ Since all the monomials $h/x_{i}$ are unique when expanding the second term of the previous equation, we deduce 
\[(P_f+P_g)\frac{h}{x_{i}}(l_{i}+l_{i}^{*})=0,\quad \forall i \in \ind(h).\]
Since none of the variables in $\ind(h/x_i)$ is present in $P_f+P_g$ or $l_{i}+l_{i}^{*}$, extracting the coefficients of $h/x_{i}$ gives 
\begin{equation}\label{eq:pr6}
   (P_f+P_g)(l_{i}+l_{i}^{*})=0,\quad \forall i \in \ind(h).
\end{equation}

Let $x_j$ be the maximum variable of $l_i+l_i^{*}.$ If $f$ and $g$ do not contain $x_j$ then we deduce that $P+f+P_g=0.$ The same outcome can be deduced if $x_j$ belongs to one of of the monomials $f,g$ or if $f=g.$    

To demonstrate that there are no collisions for addition, more exactly that $\left|\Alow_f\cdot \frac{f}{h}+\Alow_g^{\frac{f}{h}}\cdot \frac{g}{h}\right|=\left|\Alow_f\cdot \frac{f}{h}\right|\left|\Alow_g^{\frac{f}{h}}\cdot \frac{g}{h}\right|$ one can use the same techniques as in Theorem \ref{thm:Minkowski-sum}. 
\end{IEEEproof}
\section{Proof of Proposition \ref{pr:TypeI-disjoint}}
\begin{IEEEproof}
    We will demonstrate that the orbits of any of the three sub-cases do not intersect. 
\textbf{Sub-case A1.}\\
\emph{\textbf{Disjoint orbits with sub-case A1}}
Let $(f,g)\neq (f^{*},g^{*})$ with $f,g,f^{*},g^{*}\in \I_r$ and $h=\gcd(f,g),h^{*}=\gcd(f^{*},g^{*})$ satisfying $\deg(h)=\deg(h^{*})=r-\mu.$ Suppose the two orbits are not disjoint, in other words, let 
$P\in \Alow_h\cdot h\cdot (\Alow_f\cdot \frac{f}{h}+\Alow_g\cdot \frac{g}{h})$ and $P^{*}\in \Alow_h^{*}\cdot h^{*}\cdot (\Alow_f^{*}\cdot \frac{f^{*}}{h^{*}}+\Alow_g^{*}\cdot \frac{g^{*}}{h^{*}})$ be such that $P=P^{*}.$\\
If $f$ and $g$ are not comparable with respect to $\preceq$ then $f+g=f^{*}+g^{*}$, since these are the maximum monomials in $P$, respectively in $P^{*}.$ Because $f,g$ are not comparable then so is $f^{*}$ and $g^{*}$, and hence we have $f=f^{*}$ or $f=g^{*}$ which contradicts our hypothesis. \\
If $f$ and $g$ are comparable, then suppose $g\preceq f.$ Hence there is only one maximum monomial in $P$, which means that so should be for $P^{*}.$ In other words we have $f=f^{*}.$ Since $g\neq g^{*}$ we can put $i_g=\max(\ind(g/\gcd(g,g^{*})))$, the index of the maximum variable in $g$ which is not in $g^{*}.$ Do the same for $g^{*}$, i.e., $i_{g^{*}}=\max(\ind(g^{*}/\gcd(g,g^{*}))).$ We can suppose without loss of generality that $i_g>i_{g^{*}}.$ If $g$ belongs to the monomials in $P$, then belongs to the second term $\Alow_g\cdot g$ and not in $\Alow_f\cdot f.$ Since $P=P^{*}$, $g$ should also belong to the monomials in $P^{*}.$ Since the variable $x_{i_g}$ does not belong to $g^{*}$ and is strictly greater than all variables outside $\gcd(g,g^{*})$ this means that $g$ can not belong to the monomials of $\Alow_g\cdot g$ and thus has to be in the first orbit of $P^{*}$, more exactly in $\Alow_f^{*}\cdot f^{*}.$ Recall that $\gcd(f,g)=h$ with $\deg(h)=r-\mu.$ Let $h=x_{i_1}\dots x_{i_{r-\mu}}$ and $f=hx_{j_1}\dots x_{j_\mu},g=x_{l_1}\dots x_{l_{\mu}}.$ Since $f=f^{*}$ this implies that $x_{l_1}f/x_{j_1},\dots, x_{l_{\mu}}f/x_{j_{\mu}}$ belong to $P^{*}.$ However, all these monomials are greater than $g$ with respect to $\preceq$, therefore, these monomials should also belong to $P.$ This means that $h(x_{j_1}+x_{l_1})\dots (x_{j_{\mu}}+x_{l_{\mu}})$ belongs to $P$, which implies that $g$ belongs to the first term in $P$, more exactly in $\Alow_f\cdot f$ which is impossible. The same arguments apply if we suppose that $g$ does not belong to $P$, in other words $g$ belongs to both terms in $P$, i.e, to $\Alow_f\cdot f$ and $\Alow_g\cdot g.$      

\emph{\textbf{Disjoint orbits with sub-case A2}}
Suppose $(f,g)$ are given as in the previous case and that we have $f^{*},g^{*}\in\I_r$ with $h^{*}|\gcd(f^{*},g^{*})$, more exactly $\gcd(f^{*},g^{*})=h^{*}h^{'}$ with $h^{'}\neq 1.$ Let $P^{*}\in\Alow_{h^{*}}\cdot h^{*}\cdot \left(\Alow_f^{*}\cdot \frac{f^{*}}{h^{*}}+\Alow_{g^{*}}^{\frac{f^{*}}{h^{*}}}\cdot \frac{g^{*}}{h^{*}}\right).$\\
If $f,g$ are not comparable and since these are the maximum monomials in $P$, then so should we have for $P^{*}.$ This is impossible since the maximum monomials in $P$ are either both $f^{*}$ and $g^{*}$, and $\deg(\gcd(f^{*},g^{*}))>r-\mu$ (contradiction), or just $f^{*}$ ($g^{*}\preceq f^{*}$) (contradiction). \\
If $f,g$ are comparable. Suppose $g\preceq f.$ Then we have $f=f^{*}.$ Since $g\neq g^{*}$ then apply the same arguments as in the previous case: there is at least one variable on which $g$ and $g^{*}$ are different, and thus having $g$ as a monomial in $P$, from the second term ($\Alow_g\cdot g$) would mean that is also belongs to $P^{*}$, more precisely to the first term $\Alow_{f^{*}}\cdot {f^{*}}.$ This will imply using the same argument as before that $g$ will also be present in the first term of $P$, namely in $\Alow_f\cdot f$ which contradicts our assumption.    

\emph{\textbf{Disjoint orbits with sub-case B}}
Consider the same pair $(f,g)$ and a monomial $f^{*}\not\in \I_r$ satisfying the conditions from sub-case B. Suppose $P^{*}\in \Alow_{f^{*}}\cdot f^{*}+\Alow_{f^{*}}\cdot f^{*}$ such that $P=P^{*}.$ Since in $P$ either both monomials $f$ and $g$ or just one  of them is the maximum monomial with respect to $\preceq$ then these/this monomial/s should also belong to $P^{*}.$ Therefore, $f$ has to be in $P^{*}$, and on top of that there are no monomials in $P^{*}$ greater than $f$ with respect to $\preceq.$ Notice that we need to have $f\preceq f^{*}.$ By definition $P^{*}=\prod\limits_{i\in\ind(h^{*})}\left(\prod\limits_{i\in \ind(f^{*}/h^{*})}(x_i+l_i)+\prod\limits_{i\in \ind(f^{*}/h^{*})}(x_i+l_i^{*})\right).$ This means that $h^{*}| f$ and $h^{*}|f^{*}$, and therefore, we can put $f=h^{*}x_{i_1}\dots x_{i_{\mu}},f^{*}=h^{*}x_{i_1+\theta_1}\dots x_{i_{\mu}+\theta_{\mu}}.$ Recall that any monomial greater than $f$ should not belong to $P^{*}$. This implies that the restriction of $l_i$ to the set of variables that generate such monomials is equal to its corresponding $l_i^{*}.$ Without loss of generality we can set these restrictions to be equal to zero (it means that all $b_{i,j}$ from these forms are equal to zero). However, in order to generate $f$ we need to set $b_{i,j}=1$ in $l_i$ and $b_{i,j}^{*}=0$ in $l_i^{*}$ for all pairs $(i,j)\in \{(i_1+\theta_1,i_1),\dots,(i_{\mu}+\theta_{\mu},i_{\mu})$. Then the polynomial 
\[h\left(\prod\limits_{j\in[1,\mu]}(x_{i_j+\theta_j}+x_{i_j}+l_j^{\prime})+\prod\limits_{j\in[1,\mu]}(x_{i_j+\theta_j}+l_j^{*\prime})\right)\]
is a term of $P^{*}$, where $l_j^{*\prime}, l_j^{\prime}$ contain variables smaller than $x_{i_j}.$ However, this means that monomials of the form $x_{i_j}f^{*}/x_{i_j+\theta_j}$ for any $j\in [1,\mu]$ are in $P^{*}$, while being all greater than $f$, which is impossible.  

\textbf{Sub-case A2.}\\
\emph{\textbf{Disjoint orbits with sub-case A2}} 
Suppose $f,g,f^{*},g^{*}\in\I_r$ with $h^{*}|\gcd(f^{*},g^{*}), h|\gcd(f,g), $, more exactly $\gcd(f,g)=hh^{\prime}\gcd(f^{*},g^{*})=h^{*}h^{*\prime}$,  with $h^{\prime},h^{*\prime}\neq 1.$ Let $P,P^{*}$ in the orbits defined by $(f,g)$, and $(f^{*},g^{*})$, respectively. 
If $f,g$ are not comparable then the two maximum monomials in $P$ should coincide with the two maximum monomials in $P^{*}$, which yields $f=f^{*}$ or $f=g^{*}$, and ends this case.\\
If $g\preceq f$ then so should be for $P^{*}$, i.e., $g^{*}\preceq f^{*}.$ In other words $f=f^{*}.$ Since $g\neq g^{*}$ then we can apply the same technique as in the the sub-case A1 vs. A2, or A1 vs A1. \\
\emph{\textbf{Disjoint orbits with sub-case B}}
In this case consider $(f,g)$ a valid pair of monomials for the sub-case A2, and $f^{*}$ a valid monomial for the sub-case B. Since $f$ is a maximum monomial in $P$ then it should also exist in $P^{*}.$ Using the same arguments as in the sub-case A1 vs. sub-case B we have that $f\preceq f^{*}$ and when creating $f$ in $P^{*}$, we will necessary create some monomials greater than $f$ that belong to $P^{*}$, which is a impossible.  

\textbf{Sub-case B.}\\
\emph{\textbf{Disjoint orbits with sub-case B}}
Let $f\neq f^{*}$ be two valid monomials for the sub-case B. These monomials will define two polynomials (the Minkowski sum of their orbits), $P$ and $P^{*}$ that we will suppose to be equal. We will demonstrate that this is impossible. By definition we have 
\begin{align*}
    P&=\sum\limits_{i\in\ind(f)} \frac{f}{x_{i}}(l_{i}+l_i^{\prime})
        +\dots  +\prod\limits_{i\in\ind(f)}l_{i}\prod\limits_{i\in\ind(f)}l_{i}^{\prime}\\
        P^{*}&=\sum\limits_{i\in\ind(f^{*})} \frac{f^{*}}{x_{i}}(l_{i}^{*}+l_i^{*\prime})
        +\dots  +\prod\limits_{i\in\ind(f^{*})}l_{i}^{*}\prod\limits_{i\in\ind(f^{*})}l_{i}^{*\prime}
\end{align*} 

Notice that all the monomials $f/x_{i}$ for $i \in \ind(f)$ and $f^{*}/x_i$ for $i\in \ind(f^{*})$ are unique by definition of the $\Alow.$ Let us consider that $\gcd(f,f^{*})=1.$ If this is not the case we can factor by the common variables using a similar technique as in Theorem \ref{thm:Minkowski-sum}. Thus, we must have $l_i=l_i^{\prime}$ and $l_i^{*}=l_{i}^{*\prime}$ for all indices $i$ corresponding to maximum monomials. Also, due to the fact that $\gcd(f,f^{*})=1$ the maximum monomials from  $(l_i+l_i^{\prime})f/x_i$ share at most one variable with the maximum monomial from $(l_j^{*}+l_j^{*\prime})f^*/x_j$ for all indices $i,j.$ This means that all linear terms will cancel each other, i.e., $l_i=l_i^{*}$ for all $i\in\ind(f)$ and $l_j=l_j^{*}$ for all $j\in\ind(f^{*})$, leading to $P=P^{*}=0$ which is impossible.  
\end{IEEEproof}

\end{document}